\documentclass[acmsmall,screen]{acmart}
\UseRawInputEncoding % ATTENTION: comment it if raise error

\AtBeginDocument{%
  \providecommand\BibTeX{{%
    \normalfont B\kern-0.5em{\scshape i\kern-0.25em b}\kern-0.8em\TeX}}}

\usepackage{comment}
\usepackage{color}
\usepackage{url}
\usepackage{subfigure}
\usepackage{multicol}
\usepackage{xspace}
\usepackage{enumitem} % for itemize
\usepackage{diagbox}
\usepackage{multirow}
\usepackage{hyperref}
\usepackage{cleveref}[2012/02/15]% v0.18.4;

\crefformat{footnote}{#2\footnotemark[#1]#3}

\hypersetup{urlcolor=blue}

\DeclareMathOperator*{\argmin}{argmin}

\newcommand*{\rom}[1]{\uppercase\expandafter{\romannumeral #1\relax}}

\newcommand{\website}{\textcolor{blue}{\url{https://github.com/chenjshnn/WAE}}}

\setcopyright{acmcopyright}
\acmJournal{TOSEM}
\acmYear{2020}
\acmVolume{29}
\acmNumber{3}
\acmArticle{19}
\acmMonth{5}
\acmPrice{15.00}
\acmDOI{10.1145/3391613}

\begin{document}

\title{Wireframe-Based UI Design Search Through Image Autoencoder}

\author{Jieshan Chen}
\email{Jieshan.Chen@anu.edu.au}
\affiliation{%
  \institution{Australian National University}
  \country{Australia}
  }

\author{Chunyang Chen}
\email{Chunyang.Chen@monash.edu}
\authornote{Corresponding author.}
\affiliation{%
  \institution{Monash University}
  \country{Australia}
}

\author{Zhenchang Xing}
\email{Zhenchang.Xing@anu.edu.au}
\affiliation{%
  \institution{Australian National University}
  \country{Australia}
}
\additionalaffiliation{%
 \institution{Data61, CSIRO}
 \country{Australia}
}

\author{Xin Xia}
\email{Xin.Xia@monash.edu}
\affiliation{%
  \institution{Monash University}
  \country{Australia}
}

\author{Liming Zhu}
\email{Liming.Zhu@data61.csiro.au}
\affiliation{%
  \institution{Data61,CSIRO}
  \country{Australia}
}
\additionalaffiliation{%
    \institution{University of New South Wales}
    \country{Australia}
}

\author{John Grundy}
\email{John.Grundy@monash.edu}
\affiliation{%
  \institution{Monash University}
  \country{Australia}
}
\author{Jinshui Wang}
\email{ymkscom@gmail.com}
\authornotemark[1]
\affiliation{%
  \institution{Fujian University of Technology}
  \country{China}
}

\renewcommand{\shortauthors}{J. Chen et al.}

\begin{abstract}
UI design is an integral part of software development.
For many developers who do not have much UI design experience, exposing them to a large database of real-application UI designs can help them quickly build up a realistic understanding of the design space for a software feature and get design inspirations from existing applications.
However, existing keyword-based, image-similarity based, and component-matching based methods cannot reliably find relevant high-fidelity UI designs in a large database alike to the UI wireframe that the developers sketch, in face of the great variations in UI designs.
In this paper, we propose a deep-learning based UI design search engine to fill in the gap.
%The key innovation of our search engine is to train a wireframe image autoencoder for encoding the visual semantics of UI designs in a dense vector space, in which similar UI designs can be easily found by k-nearest neighbor search.
%The wireframe autoencoder is trained using a large database of real-application UI designs, without the need for labeling relevant UI designs.
The key innovation of our search engine is to train a wireframe image autoencoder using a large database of real-application UI designs, without the need for labeling relevant UI designs.
We implement our approach for Android UI design search, and conduct extensive experiments with artificially-created relevant UI designs and human evaluation of UI design search results.
Our experiments confirm the superior performance of our search engine over existing image-similarity or component-matching based methods, and demonstrates the usefulness of our search engine in real-world UI design tasks.
\end{abstract}

\begin{CCSXML}
<ccs2012>
<concept>
<concept_id>10011007</concept_id>
<concept_desc>Software and its engineering</concept_desc>
<concept_significance>500</concept_significance>
</concept>
<concept>
<concept_id>10011007.10011006.10011072</concept_id>
<concept_desc>Software and its engineering~Software libraries and repositories</concept_desc>
<concept_significance>500</concept_significance>
</concept>
<concept>
<concept_id>10011007.10011074.10011092.10010876</concept_id>
<concept_desc>Software and its engineering~Software prototyping</concept_desc>
<concept_significance>500</concept_significance>
</concept>
</ccs2012>
\end{CCSXML}

\ccsdesc[500]{Software and its engineering}
\ccsdesc[500]{Software and its engineering~Software libraries and repositories}
\ccsdesc[500]{Software and its engineering~Software prototyping}

\keywords{Android, UI search, Deep Learning, Auto-encoder.}

\maketitle

\section{Introduction}
\label{sec:introduction}
Graphical User Interface (GUI) is ubiquitous in modern desktop software, mobile applications and web applications.
It provides a visual interface between a software application and its end users through which they can interact with each other.
A well-designed GUI makes an application easy, practical and efficient to use, which significantly affects the success of the application and the loyalty of its users~\cite{winograd1995programming, jansen1998graphical, web:fishEar}.
For example, in the competitive mobile application market, the design of an application's GUI, or even its icon, has become crucial for distinguishing an application from competitors, attracting user downloads, reducing users' complaints and retaining users~\cite{miniukovich2016pick, doosti2018computational, web:googleTranslate}.

\begin{figure*}
	\centering
	\includegraphics[width=0.9\textwidth]{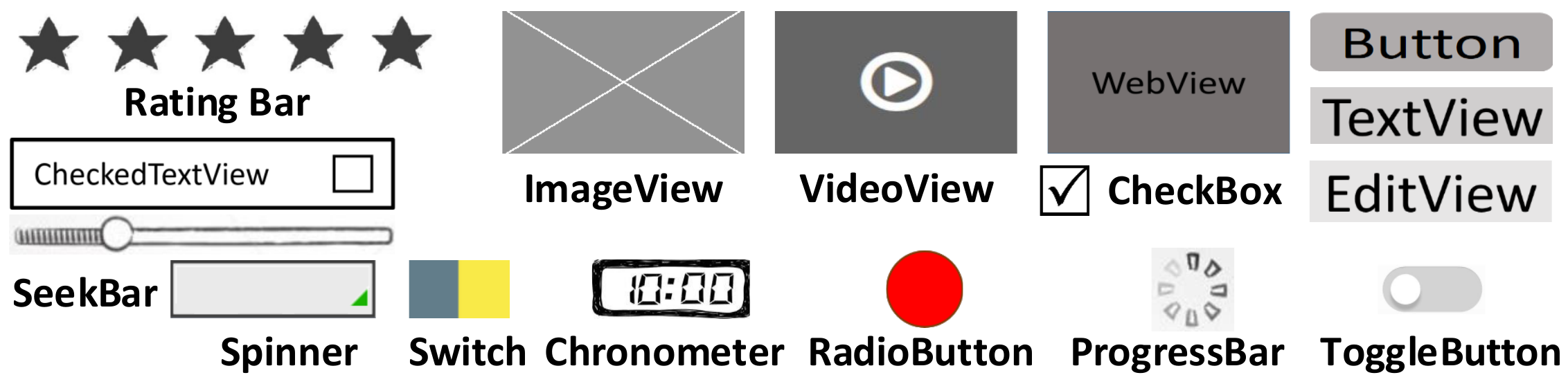}
    %\vspace{-3mm}
	\caption{The most common wireframe components for Android UI design. Each component has its own function. For example, TextView will show text to the user and EditText enables user to input text.}
    \vspace{-3mm}
	\label{fig:wiriframeComponents}
\end{figure*}

Designing the visual composition of a GUI is an integral part of software development.
Based on the initial user needs and software requirements, the designers usually first design a \textit{wireframe} of the desired GUI by selecting highly-simplified visual components with special functions (for example those shown in Figure~\ref{fig:wiriframeComponents}) and determining the layout of the selected components that can support the interactions appropriate to application data and the actions necessary to achieve the goals of users, and modify their designs iteratively by comparing with existing online design examples.
They then add high-fidelity visual effects to the GUI components, such as colors and typography, and add application-specific texts and images to the GUI design.
Of course, the wireframe design and the high-fidelity GUI design are interweaving and iterative during which designers continually explore the design space by removing unnecessary visual components, adding missing components, and refining the components' layout and visual effects.

To satisfy users' needs, designing a good GUI demands not only the specific knowledge of design principles and guidelines (e.g., Android Material Design~\cite{web:materialDesign}, iOS Human Interface Guidelines~\cite{web:iosDesign}), but also the understanding of design space which has the great variations in visual components that can be potentially used, their layout options, and visual effect choices.
As shown in Figure~\ref{fig:storyLine}, the design space of a GUI, even for the simple sign-up feature, can be very large.
However, due to the shortage of UI designers~\cite{hong2011matters}, software developers who do not have much understanding of UI design space often have to play the designer role in software development, especially in the start-up companies and open-source projects.
For example, in a survey of more than 5,700 developers~\cite{web:developerDesign}, 51\% respondents reported that they do not have much UI design training but they work on UI design tasks, more so than other development tasks.
In fact, when developing an application, what software developers and designers focus on are totally different.
Developers try to make the application work while designers target at making it adorable~\cite{web:whyUIDesignHard}, which makes it tough for software developers to directly work as designers.
An effective mechanism is needed to support such developers to explore and learn about the UI design space in their UI design work.

Providing developers with a UI design search engine to search existing UI designs can help developers quickly build up a realistic understanding of the design space of a GUI and get inspirations from existing applications for their own application's UI design.
However, compared with the well-supported code search~\cite{paul1994framework, wang2013improving, peng2011iterative, mcmillan2011portfolio}, there has been little support for UI design search.
Existing UI design search methods~\cite{web:dribbble, web:UImovement, bernal2019guigle} are based on keywords describing software features, UI design patterns or GUI components.
Although keyword-based UI search could provide some initial design inspirations, a more advanced UI design search engine is still needed to explore the design space in a more targeted manner, which can directly take as input a schematic UI (e.g., a wireframe) that the developers sketch and returns high-fidelity UI designs alike to the input (see Section~\ref{sec:motivation} for a motivating scenario).
However, a few keywords can hardly describe the visual semantics of a desired UI design, such as visual components used and their layout.

As the advanced UI design search can consider UI designs as images, a naive solution could be searching UI designs by image similarity of certain image features such as color histogram~\cite{jain1996image} or Scale-Invariant Feature Transform (SIFT)~\cite{lowe1999object}.
Although such image features are useful for measuring image similarity, they are agonistic of the visual semantics (visual components and their compositions) of a GUI.
As such, image-wise similar UI designs are very likely design-wise irrelevant.
Alternatively, one can heuristically match individual visual components based on their type, position and size for measuring the similarity of two UI designs~\cite{reiss2018seeking, behrang2018guifetch, zheng2019faceoff}.
However, such methods are restricted by the pre-defined component matching rules, and is sensitive to the cut-off matching thresholds.
Furthermore, individual component-matching heuristics often retrieve many irrelevant UI designs, because individual component matching cannot effectively encode the visual composition of components in a GUI as a whole.

The most related work is Rico~\cite{deka2017rico}, which introduces a new UI dataset, discusses five potential usages and demonstrates the possibility of assisting UI search.
Their dataset is collected by automatic app exploration and manual exploration by recruiting crowd workers.
In terms of UI search demonstration, they use a simple multilayer perceptron with only six fully connected layers, simplify the UI components as text/non-text, and show several examples without any detailed statistics on performance results.
As discussed above, UI designs are sophisticated with many variants, so it is impossible to merely use text and non-text to express the core concepts of one UI design.
Besides, the complex combination of different widgets with arbitrary numbers and positions shows a huge design space in terms of typology.
Therefore, a naive method with highly simplified widgets is not enough to tackle this task.
Note that the Rico evaluation only shows several examples without any detailed studies on retrieval accuracy, data issue, model limitation, failure cases and usefulness evaluation.
Thus, we cannot know the generalization or performance of their model.

In this paper, we present an approach to develop a deep-learning-based UI design search engine using a convolutional neural network instead of purely fully connected layers.
To expose developers to diverse, real-application UI designs for a variety of software features, we use reverse-engineering method (such as the automatic GUI exploration methods in~\cite{chen2018ui}) to build a large database of UI screenshots (and their corresponding wireframes) of existing applications.
We further identify 16 user interaction components which narrow the gap between designers and developers by analyzing several design platforms and UI implementation details.
Our approach performs wireframe-based UI design search.
A wireframe captures the type and layout information of visual components, but ignores their high-fidelity visual details.
As such, they can be fast prototyped and refined with minimal effort, and can retrieve visually-different but semantically relevant UI designs (see Figure~\ref{fig:storyLine} for example).
Our approach does not perform individual component matching, but it attempts to judge the relevance of the whole UI designs.
A key challenge in developing such a robust UI-design relevance model is that no labelled relevant UI designs exist and it requires heavy manual efforts to annotate such a large dataset.
Thus, we cannot use supervised learning methods like~\cite{chen2016learning, krizhevsky2012imagenet} to train the model for encoding the visual semantics of UI designs.
To overcome this challenge and relieve the heavy manual efforts, we design a wireframe autoencoder which can be trained using a large database of UI wireframes in an unsupervised way.
Once trained, this autoencoder can encode both the query wireframe by the user and the UI screenshots of existing applications through their corresponding wireframes in a vector space of UI designs.
In this vector space, retrieving UI screenshots alike to the query wireframe can be easily achieved by k-nearest neighbors (kNN) search.

As a proof of concept, we implement our approach for searching Android mobile application UI designs in a database of 54,987 UI screenshots from 25 categories \footnote{Since Google Play updated their app categories after our data collection, the number of categories (25) of our dataset is different from the current number (35)  in Google Play.} of 7,748 top-downloaded Android applications in Google Play.
We evaluate the performance, generalization and usefulness of our UI design search engine\footnote{All UI design images in this paper and all experiment data and results can be downloaded at our Github repository \website} with an automatic evaluation of 4500 pairs of relevant UI designs generated by component-scaling and component-removal operations, the human evaluation of the relevance of the top-10 UI designs returned for 50 unseen query UIs from 25 applications (not in our database), and a user study with 18 non-professional UI designers on five UI design tasks.
Our evaluation confirms the superior performance of our approach than the baselines based on low-level image features (color histogram and SIFT), individual component-matching heuristics and fully connected layers based neural network.
The user study participants highly appreciate the relevance, diversity and usefulness of UI design search results by our tool in assisting their design work.
They also point out several user needs for UI design search, such as constraint-aware UI design search, more flexible encoding of component layouts.

Our contributions can be summarized as follows:
\begin{itemize}
 \item We propose a novel deep-learning based approach using convolutional neural network in an unsupervised manner for building a UI design search engine that is flexible and robust in face of the great variations in UI designs.
 \item We build a large wireframe database of UI designs of top-downloaded Android applications by exploring different wireframing approaches, and develop a web-based search interface to implement our approach.
 \item Our extensive experiments demonstrate the performance, generalization and usefulness of our approach and tool support, and point out interesting future work.
\end{itemize}

The rest of the article is organized as follows.
Section~\ref{sec:motivation} presents a motivating scenario to describe the potential usage of our tool.
Section~\ref{sec:approach} describes the general approach of data collection, model structure and the principle of our UI design search engine.
Section~\ref{sec:Implementation} reveals the detailed methodology of the selection of types and representations of wireframes, the selection of model hyperparameters and the tool implementation.
We describe our experiment setup and results in Section~\ref{sec:experiments_setup} and Section~\ref{sec:results} respectively, in terms of the best representation of wireframes, accuracy, generalization and usefulness.
Potential threats to validity are discussed in Section~\ref{sec:threatsToValidity}.
Section~\ref{sec:relatedWork} discusses related work and Section~\ref{sec:conclusion} concludes the article.

\section{Motivating Scenario}
\label{sec:motivation}

\begin{figure*}
	\centering
	\includegraphics[width=0.9\textwidth]{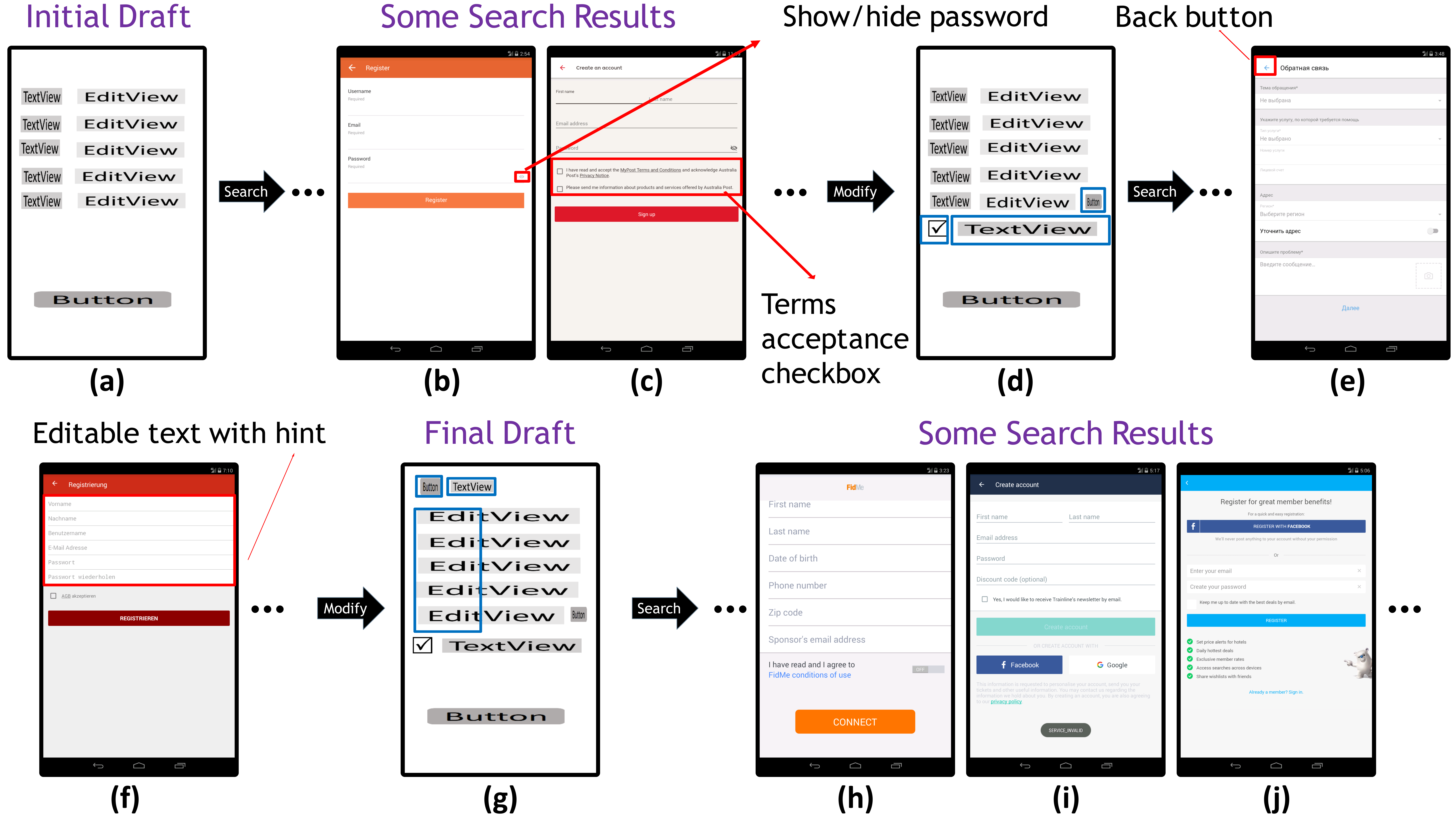}
    %\vspace{-7mm}
	\caption{UI design search: benefits and challenge}
	\label{fig:storyLine}
\end{figure*}

A start-up company needs to design the UIs for its mobile application.
Like many small companies~\cite{hong2011matters}, it does not have a professional UI designer due to budget constraints.
So the design work is assigned to a software developer Lucy.
Lucy has some desktop software front-end development experience, but never designs a mobile application UI.

The first task for Lucy is to design a sign-up UI for collecting user information, such as user name, password and email, during user registration.
Based on her prior desktop software development experience, Lucy designs a very basic sign-up UI (Figure~\ref{fig:storyLine} (a)).
It has several side-by-side TextView and EditView: TextView for displaying a label for the information to be collected, and EditView for entering the information.
At the bottom, it has a button for submitting the entered user information.

Lucy is afraid that her design is not complete, nor trendy for mobile applications.
She would like to see if other applications design sign-up UIs like hers, but she does not want to randomly download and install applications from app market just to see their sign-up UIs (if any).
Not only is it time-consuming, but it also cannot give a systematic view of relevant UI designs.
A better solution is to feed her UI design into an effective UI design search engine which can return similar but visual-effect-diverse UI designs from a large database of UI designs.

Lucy tries to use such a UI design search engine to obtain a list of UI designs alike to her initial sign-up UI design.
Observing the returned UI designs, Lucy realizes that although her initial design has the basic functionality, it does miss some nice and important features.
For example, she can add a show/hide password button (e.g., Figure~\ref{fig:storyLine} (b)), which is convenient for users to confirm the entered password.
Furthermore, sign-up UI is a good place for users to access and acknowledge relevant terms and conditions (e.g., Figure~\ref{fig:storyLine} (c)).
Based on such observations, Lucy refines her design as the one in Figure~\ref{fig:storyLine} (d) (changes highlighted in blue box) and search the UI design database again.

Observing the search results, Lucy gains a realistic understanding of what a trendy sign-up form needs, including visual components, layout options and visual effects, and further refines her design.
%further realizes that her design is not trendy for mobile applications.
For example, mobile applications often have a navigation button at the top (e.g., the back button in Figure~\ref{fig:storyLine} (b)/(c)/(e)/(f)) to facilitate the navigation among UI pages.
Furthermore, unlike the traditional side-by-side label-text input design in desktop software, mobile applications use an editable text with hint to achieve the same effect (Figure~\ref{fig:storyLine} (c)/(e)/(f)).
This design works better for mobile devices which have much smaller screens than desktop computer.
Based on these design inspirations, Lucy further refines her design as the one in Figure~\ref{fig:storyLine} (g).
Comparing the UI design search results ((e.g., Figure~\ref{fig:storyLine} (h)/(i)/(j)) with the design in Figure~\ref{fig:storyLine} (g), Lucy is now confident in her final UI wireframe.
Furthermore, observing many relevant and diverse UI designs gives Lucy many inspirations for designing high-fidelity visual effects (e.g., color system, typography) for her UIs.

As UI designs in Figure~\ref{fig:storyLine} shows, the design space of a GUI can be very huge, with the great variations in:
(1) the type of visual component used (e.g., checkbox in many UI designs versus switch unique in Figure~\ref{fig:storyLine} (h));
(2) the number of visual components in a design (e.g., editable text and button in Figure~\ref{fig:storyLine} (g), (i) and (j));
(3) the position and size of visual components (e.g.,  editable text and button in Figure~\ref{fig:storyLine} (b) versus (h));
and (4) the layout of visual components (e.g., side-by-side label-textinput in Figure~\ref{fig:storyLine} (a) versus up-down label-textinput in Figure~\ref{fig:storyLine} (b), or left-checkbox + right-text in Figure~\ref{fig:storyLine} (c) versus left-text + right-switch in Figure~\ref{fig:storyLine} (h)).
Achieving the above-envisioned benefits of UI design search requires the search engine to be flexible and robust in face of the great variations, and to achieve a good balance between similarity and variation in UI designs.

\begin{figure*}
	\centering
	\includegraphics[width=1.0\textwidth]{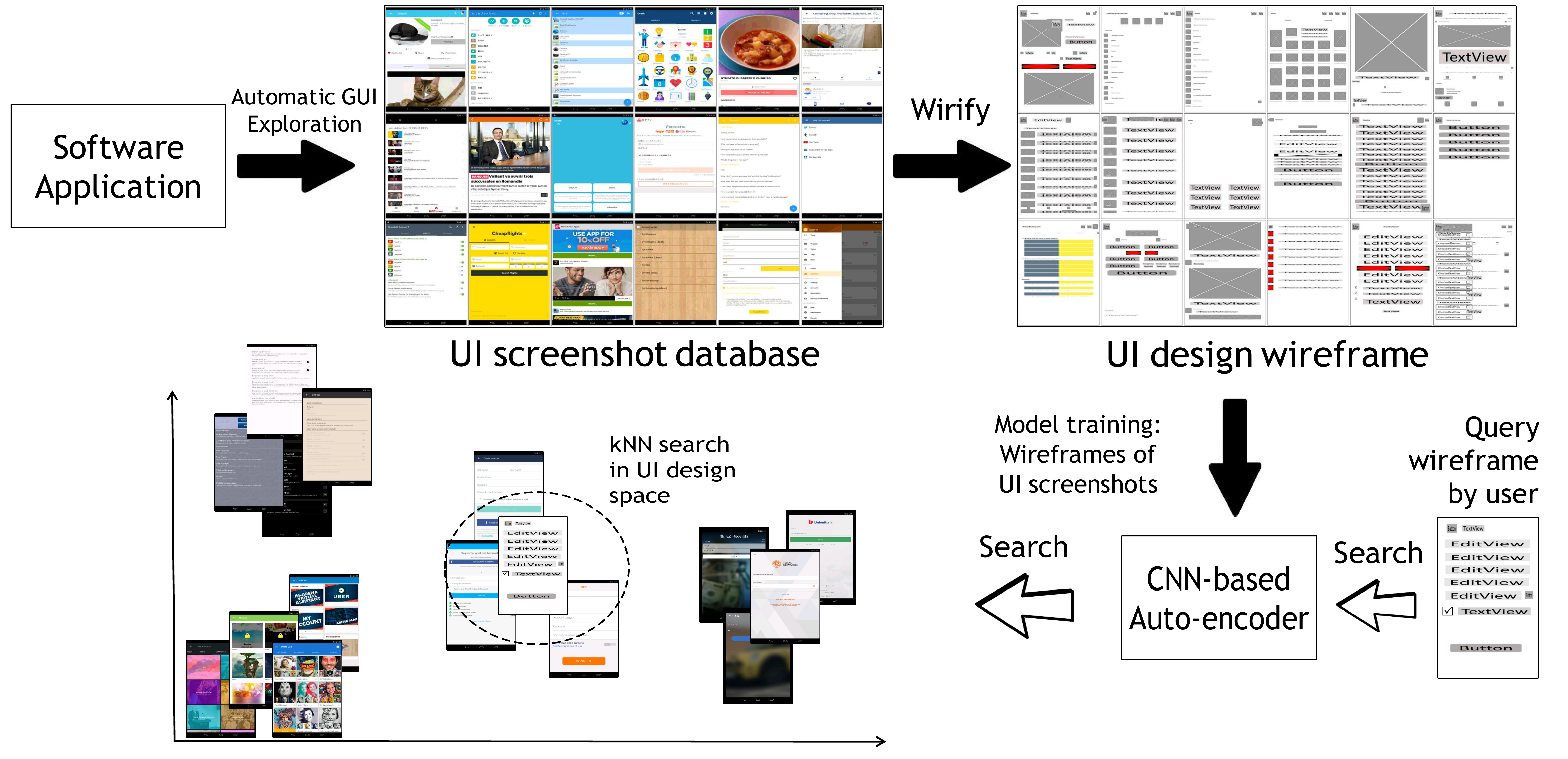}%
	\caption{An overview of our approach}
	\label{fig:flowChart}
\end{figure*}

\section{Our Approach}
\label{sec:approach}
Figure~\ref{fig:flowChart} presents the overview of our deep-learning based approach for building a UI design search engine that is flexible and robust in the face of the great variations in UI designs.
Our approach consists of three main steps:
1) build a large database of diverse, real-application UI designs using automatic GUI exploration based methods (Section~\ref{sec:uidesigndatabase});
2) train a CNN-based wireframe autoencoder for encoding the visual semantics of UI designs using a large database of UI design wireframes (Section~\ref{sec:autoencoder}); and
3) embed the UI designs in a latent vector space using the trained wireframe encoder and support wireframe-based kNN UI design search (Section~\ref{sec:knn})

%\section{Data}

\subsection{Large Database of Real-Application UI Designs}
\label{sec:uidesigndatabase}
A large database of diverse, real-application UI designs for a variety of different software features is necessary to expose developers to the realistic UI design space.
To that end, we adopt automatic data collection method to first build a large database of UI designs from existing applications, {and then use collected data to further construct our wireframe dataset.

\subsubsection{Automatic GUI Exploration}
\label{sec:reverse_engineering}
Different techniques can be used for automatically explore the GUIs of mobile applications~\cite{chen2018ui, deka2017rico}, web applications~\cite{kumar2013webzeitgeist, ritchie2011d}, or desktop applications~\cite{bao2015tracking}.
Although technical details are different, these techniques work conceptually in the same way.
They automatically explore the GUI of an application by simulating user interactions with the application, and output the GUI screenshot images and the runtime visual component information which identifies each component's type and coordinates in the screenshots.
During the GUI exploration process, the same GUIs may be repeatedly visited, but the duplicated screenshots are discarded to ensure the diversity of the collected UI designs.
To enhance the quality of the collected UI designs, further heuristics can be implemented to filter out meaningless UIs, for example, the home screen of mobile device, the simple UIs without much design like a UI with only one image component.
In detail, we first crawl apps from Google Play, and automatically install and run the app in the simulator.
For each app, our simulator interacts with the app by simulating the user's actions including clicking buttons, entering text, and scrolling the screen.
When entering one new page, our tool will take a screenshot of the current UI and dump the XML runtime code.
The XML runtime code contains all information about the current UI, including all contained components with their corresponding bounds, class, text, boolean attributes regarding executability (such as checkable, clickable and scrollable) and the hierarchical relationship among them.
To ensure the coverage of our explored UIs, we also apply the rules set in ~\cite{chen2018ui}, which define the probability (or weight) of each potential actionable component to be pressed.
There rules are defined as:
(1) actions with higher frequency are given lower weights since we need to give other rare actions chance to perform;
(2) actions which would lead to more subsequent UIs would have higher weights in order to explore various UIs;
and (3) some special actions (such as \textit{hardware back} and \textit{scroll}) would be controlled in case they close current page or impact others' actions at the wrong time.
The actual weights of each executable components are given by
$ \textit{weights}(a) = ( \alpha * T_{a} + \beta * C_{\alpha}) / \gamma * F_{a} $,
where $a$, $T_{a}$, $C_{\alpha}$ are the action, the weights of different types of actions and the number of unexplored executable components in current UI respectively, and $\alpha, \beta, \gamma$ are the hyperparameters.
Since this collection process is automatic, some UIs may be revisited several times and we need to remove duplicate data.
To this end, we compare current dumped XML code files with the collected data by comparing the hash value of GUI component sequences.

\subsubsection{Wirification}
\label{sec:wirification}
Our approach performs wireframe-based UI design search.
Therefore, different from existing reverse-engineering methods, we need to further obtain a UI wireframe for each collected UI screenshot in the database.
To that end, we have two steps.
First, we define a set of wireframe components that are essential for different kinds of user interactions at the design level by analysing popular designer's tools and the underlying implementation details of UIs.
Second, we find the ``right'' representation of each component by our exploration experiments.

\begin{figure}
	\centering
	\includegraphics[width=0.8\textwidth]{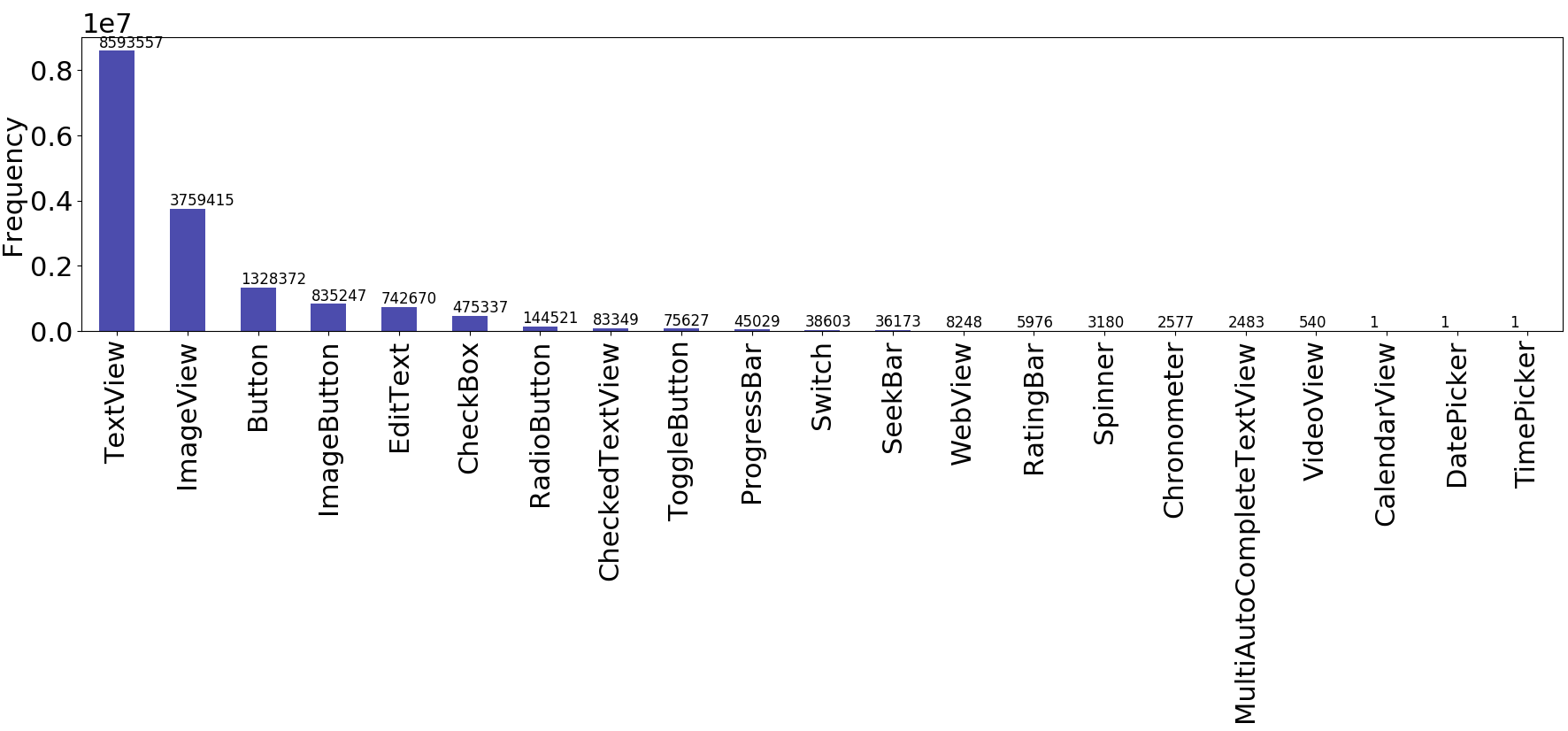}
	\caption{Bar chart of the frequency of each component contained in our dataset}
	\label{fig:frequncy_widgets}
\end{figure}

\textbf{Selection of component types.}
First, there are two types of Android UI components in terms of functionality: layout components and UI control components~\cite{android-ui}.
The UI control components (e.g., button, textView, ImageView) are the visible components we could see and interact with, while layout components (e.g., linearLayout, relativeLayout) are used for constraining the position relationship among UI control components.
As the input of this work is just the wireframe which involves more about the control component selection with rough position, we are concerned with only UI control components (we briefly mention it as UI components).
There are 21 UI components in our dataset, which we choose as the candidate components for the wireframe.
When converting an UI screenshot with corresponding run-time code, we consider two factors:
(1) Components with similar function and similar visual effect would not have much difference when designing the wireframe;
(2) Components which are rarely used may not be very useful for the UI design.
For the first consideration, we merge MultiAutoCompleteTextView with EditText.
Both of them enable editable text, but MultiAutoCompleteTextView has additional text auto-complete function.
However, this function can be achieved in EditText by manifesting the underlying background code.
We also merge ImageButton with Button because they both enable users to click them and then trigger some events.
In terms of the second consideration, we ignore CalenderView, TimePicker and DatePicker components as they appear only once in our dataset(see Figure~\ref{fig:frequncy_widgets}).
The low frequency may because they further separate into several children components, such as TextView and Spinner.
As a result, we leave with 16 components as our final set of wireframe units.
The 16 types of components are also widely covered by popular wireframe tools for mobile UI design like Adobe XD~\cite{AdobeXd}, Fluid UI~\cite{FluidUI}, Balsamiq Mockups~\cite{BalsamiqMockups}.
In the implementation of the wireframing process, we use the representation of EditText to represent MultiAutoCompleteTextView, and the representation of Button to represent ImageButton in the wireframe.
We do not draw CalenderView, TimePicker and DatePicker in the wireframe for the above reason.
For other components, we draw them with their own representations in the wireframe.
We release the source code of the wireframe transformation in our Website\footnote{\website}.

\begin{figure}
	\centering
	\includegraphics[width=0.7\textwidth]{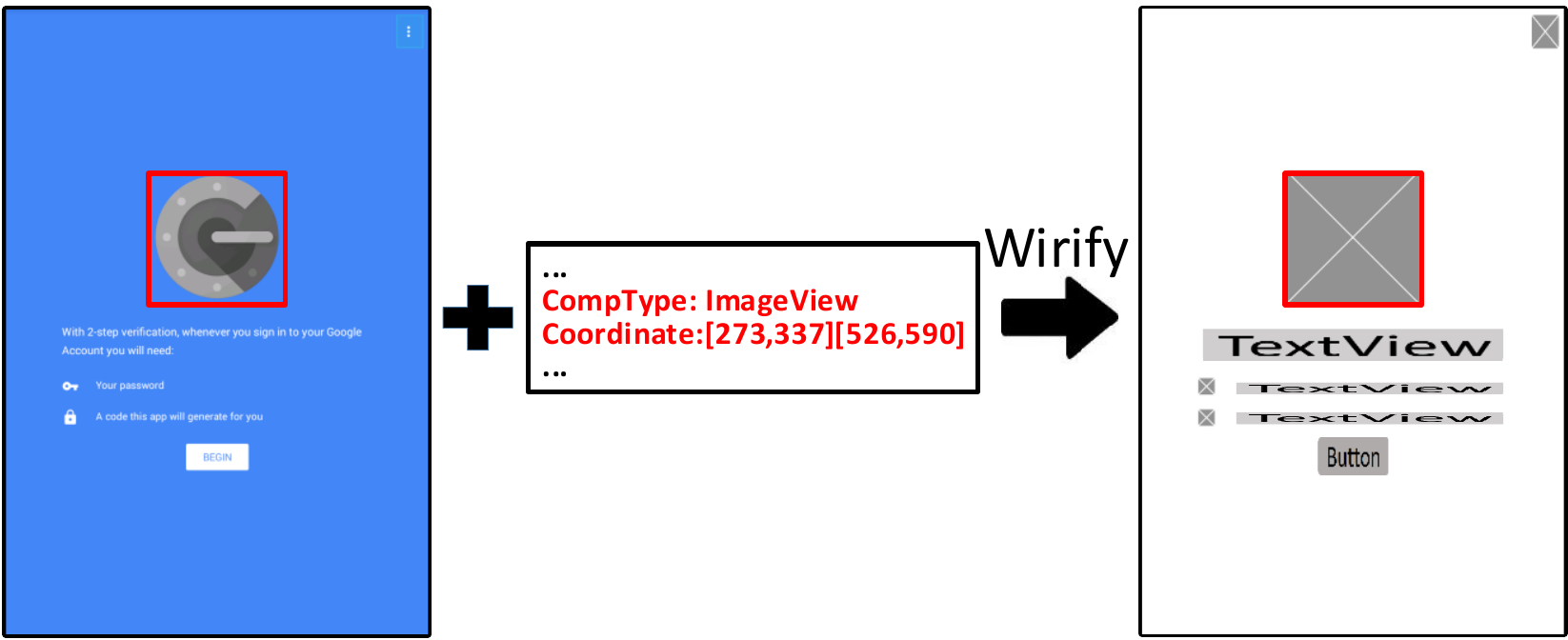}
	\caption{The visual rendering of the wireframe illustrates that top large ImageView.}
	\label{fig:visualRendering}
\end{figure}

\textbf{Wirification Process.}
After defining the 16 core wireframe components, a UI screenshot is then wirified into a UI wireframe using the XML runtime code file we dumped during the automatic explorations of apps.
Note that there is no uncertainty during this process as it is completely a rule-based process.
We wireframe the screenshot according to its dumped run-time code directly from the Android operating system, which contains the type and coordinates of each component in a UI screenshot.
Therefore, these UI screenshots and the corresponding runtime code files are perfectly matched, and there will be no error during the wireframe transformation.
Figure~\ref{fig:visualRendering} illustrates this high-level wirification process:
the UI wireframe is of the same size as the UI screenshot, and has a white canvas on which a wireframe component is drawn at the same position and of the same size as each corresponding visual component in the UI screenshot (e.g., ImageView).
However, the wireframe components ignore the color and the text/image content of the corresponding visual components.

\textbf{Exploration of the best representation way of wireframes.}
In addition to this, we need to define the representation of these components to construct our final wireframe dataset.
We do not use the default images of popular tools~\cite{AdobeXd,BalsamiqMockups, FluidUI} as they are not precise or general enough.
Instead, we represent them with simple rectangles in different colors, which can explicitly tell the model that those components are different.
Due to the huge design space, it is unrealistically to consider all colors and color palettes, so we consider three typical variants to represent these visual components, namely different grey-scale values, different colors, and different colors with different textures.
The detailed exploration setup and results of the best representation of components will be discussed later in Section~\ref{sec:ChoiceOfColor} and Section~\ref{sec:rq1_results} respectively.
Note that to avoid potential distraction, we present our wireframe as text with different grey-scale color background to help readers better understand these wireframes in the paper.

\subsection{CNN-Based Wireframe Autoencoder}
\label{sec:autoencoder}
Determining the relevance of UI designs is a challenging task, in that it requires encoding not only visual components individually, but also the visual composition of the components in a UI as a whole.
The design space of what components to use and how to compose them in a UI is huge, and thus cannot be heuristically enumerated.
CNN-based model can automatically learn latent features from a large image database, which has outperformed hand-crafted features in many computer vision tasks~\cite{lecun2015deep, krizhevsky2012imagenet, girshick2015fast}.
Although we have a large database of UI designs, the relevance of these UI designs are unknown.
Therefore, we have to train a CNN model for encoding the visual semantics of UI designs in an unsupervised way.
To that end, we choose to use a CNN-based image autoencoder architecture~\cite{vincent2010stacked} that requires only a set of unlabeled input images for model training.
As illustrated in Figure~\ref{fig:autoencoder}, our autoencoder takes as input a UI wireframe image.
It has two components: an encoder compresses the input wireframe into a latent vector representation through convolution and downsampling layers, and then a decoder reconstructs an output image from this latent vector representation through upsampling and transposed convolution layers.
The reconstructed output image should be as similar to the input image as possible, which indicates that the latent vector captures informative features from the input wireframe design for reconstructing it.
This latent vector representation of UI designs can then be used to measure the relevance of UI designs.

\begin{figure*}
	\centering
	\includegraphics[width=0.95\textwidth]{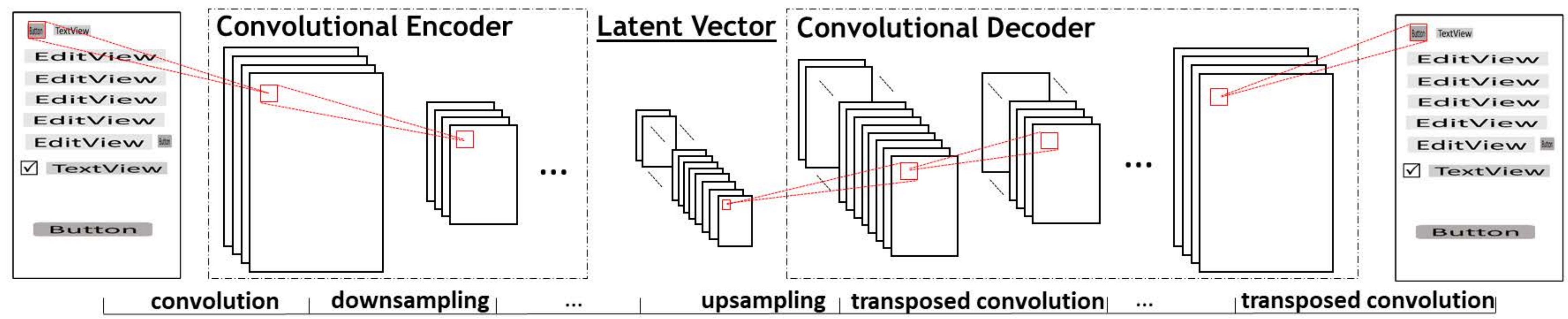}
	\caption{The architecture of our wireframe autoencoder}
	\label{fig:autoencoder}
\end{figure*}

\subsubsection{Convolution}
A convolution operation performs a linear transformation over an image such that different image features become salient.
According to the research of CNN visualization~\cite{zeiler2014visualizing, lecun2015deep}, shallow convolutional layers detects simple features such as edges, colors and shapes, which are then composed in the deep convolutional layers to detect domain-specific features (e.g., the visual semantics of UI designs in our work).

An image is represented as a matrix of pixel values, i.e., $ 0 \leqslant p_{hwd} \leqslant 255$ where $h$, $w$ and $d$ are the height, width and depth of the image.
$d=1$ for grayscale image and $d=3$ for RGB color image.
The convolution of an image uses a $kernel$ (i.e., a small matrix like $3 \times 3 \times d$ of learnable parameters) and slide the kernel over the image's height and width by 1 pixel at a time.
At each position, the convolution operation multiplies the kernel element-wise with the kernel-size subregion of the image, and sums up the values into an output value.
The transposed convolution is the opposite to the normal convolution.
It multiples a value with a kernel and outputs a kernel-size matrix.
%The output value is fed into $ReLU(x) = max(0, x)$ to perform a non-linear activation.
A convolutional layer can apply a number of kernels ($n$).
The output matrix ($h \times w \times n$) after a convolutional layer is called a feature map, which can be fed into the subsequent network layers for further processing.
Each kernel map $h \times w$ in the feature map corresponds to a kernel, and can be regarded as an image with some specific features highlighted.

\subsubsection{Downsampling \& Upsampling}
Within the encoder, downsampling (also called pooling) layers take as input the output feature map of the preceding convolutional layers and produce a spatially (height and width) reduced feature map.
A downsampling layer consists of a grid of pooling units, each summarizing a region of size $z \times z$ of the input kernel map.
As the downsampling layer operates independently on each input kernel map, the depth of the output feature map remains the same as that of the input feature map.
In our architecture, we adopt \textit{1-max pooling}~\cite{murray2014generalized} which takes the maximum value (i.e., the most salient feature) in the $z \times z$ region.
1-max pooling brings the benefits of the invariance to image shifting, rotation and scaling, leading to a certain level of insensitivity to encoding component spatial variations in UI designs.

Within the decoder, we use the upsampling layers which are opposite to downsampling.
They increase the spatial size (height and width) of the feature map by replacing each value in the input feature map with multiple values.
In our architecture, we adopt the nearest-neighbor interpolation~\cite{keys1981cubic}, i.e., enrich the original pixel in the feature map into a $z \times z$ region with the same value as the original pixel.
The upsampling layers progressively increase the spatial size of the feature map until the decoder finally reconstructs an output wireframe of the same size as the input wireframe.

\begin{figure*}
	\centering
	\includegraphics[width=0.99\textwidth]{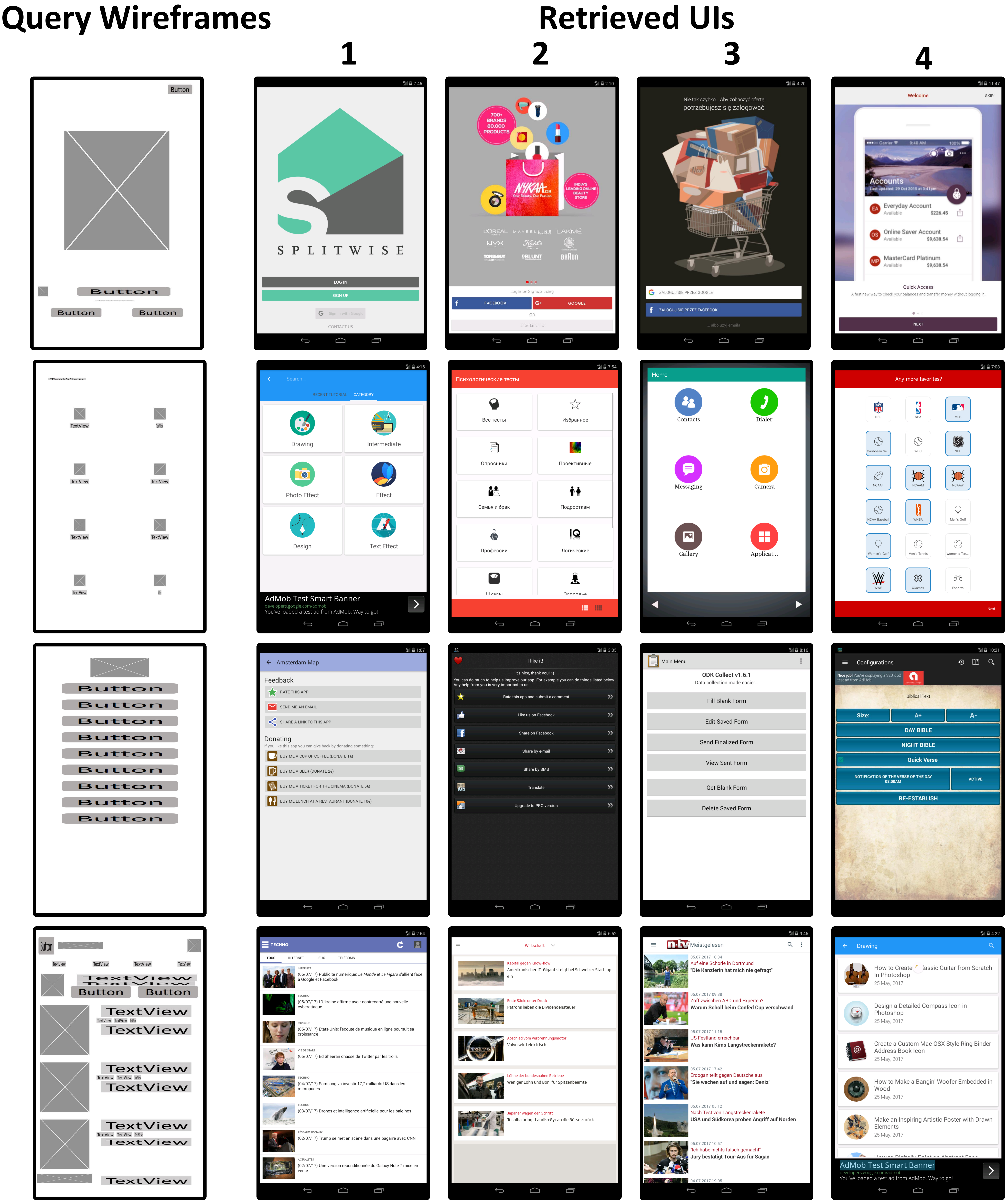}
	%\vspace{-3mm}
	\caption{Examples of kNN search in UI design space}
	\vspace{-3mm}
	\label{fig:examples}
\end{figure*}

\subsubsection{Model Training}
The encoder and the decoder are trained as an end-to-end system.
Given a UI wireframe image $X$, the encoder compresses it into the latent vector $V$: $\phi: X \rightarrow V$ where $\phi$ represents the function of the encoder's convolutional and downsampling layers.
Then the decoder decodes the latent vector $V$ into an output wireframe image $Y$:
$\psi: V \rightarrow Y$ where $\psi$ represents the function of the decoder's upsampling and transposed convolutional layers.
The target is to minimize the difference between the input wireframe $X$ and the output wireframe $Y$: $\argmin_{\phi, \psi} \| X - Y\|^2$.
We train our wireframe autoencoder to minimize the reconstruction errors with mean square error (MSE)~\cite{berger2013statistical}, i.e., $\mathcal{L}(X, Y) = \| X- Y \|^2 $.
At the training time, we optimize the MSE loss over the training dataset using stochastic gradient descent~\cite{bottou2010large}.
The decoder backpropogates error differentials to its input, i.e., the encoder, allowing us to train a wireframe encoder using unlabelled input wireframes.

\subsection{kNN Search in UI Design Space}
\label{sec:knn}
As shown in Figure~\ref{fig:storyLine}, by training the wireframe autoencoder, we obtain a convolutional encoder which can encode an input wireframe into a latent vector representation.
Given a database of automatically-collected UI screenshots (can be different from the UI screenshots used for model training), we use this trained wireframe encoder to embed the UI screenshots through their corresponding wireframes into a UI design space $S$.
Each UI screenshot $uis$ is represented as a latent vector $V(uis)$ in this UI design space.
Given a query wireframe $wf_q$ drawn by the user, we also use the trained wireframe encoder to embed $wf_q$ into a vector $V(wf_q)$ in the UI design space.
Then, we perform k-nearest neighbors (kNN) search in the UI design space to find the UI screenshots $uis$ whose embedding is the top-k most similar (by Mean Square Error (MSE) in this work) to that of the query wireframe, i.e., $\argmin_{uis \in S}^k ||V(uis), V(wf_q)||^2$.
Figure~\ref{fig:examples} shows some examples of UI design results from our empirical studies.
The fourth example shows that our model can successfully encode the visual semantics of rather complex UI designs.

\section{Proof-of-Concept Implementation}
\label{sec:Implementation}
\subsection{Data Collection}
We develop a proof-of-concept tool for searching Android mobile application UI designs.
The backend UI design space contains 54,987 UI screenshots from 7,748 Android applications belonging to 25 application categories.
We crawl the top-downloaded Android applications from Google Play, because studies show that the download number of an application correlates positively with the quality of the application's GUI design~\cite{miniukovich2016pick, doosti2018computational}.
There are three types of Android Apps: native, hybrid and web apps~\cite{web:app_type}.
The underlying implementations of these types are different.
Native apps use Android native widgets or widgets derived from them, hybrid apps utilize WebView to encode their HTML/CSS part components into an Android Application, and web apps directly use HTML/CSS/JavaScript.
In our paper, we only collect UIs from native and hybrid applications because they are easy to download and install from app store, while there is no such ``app store'' for web applications.
We remove some UIs, whose WebView takes over half of the screen.
We keep small WebView component because most of them are advertisement

We use the automatic GUI exploration method in~\cite{chen2018ui} to build a large database of UI screenshots from these Android applications, and the detailed process is stated in Section~\ref{sec:reverse_engineering}.
In total, we crawled 8,000 Android apps from Google Play with the highest installation numbers and successfully ran 7,748 Android applications and collected 54,987 UI screenshots.
Note that some apps were discarded due to the need of extra hardware support or the absence of some certain third party libraries in our emulator.
The median number of UI screenshots per application is three.
Our database contains very diverse UI designs (see Figure~\ref{fig:flowChart} for some randomly selected examples).
More examples can be seen in our Github repository\footnote{\label{web2}\website}.
%We can observe the diversity of UI designs in our database.

\subsection{Model Hyperparameters}
The wireframe autoencoder in our tool is configured as follows.
The input wireframe is a RGB color image and scaled to $180 \times 228$ for efficient processing.
The encoder uses four convolutional layers, which use 16 3x3x3 kernels, 32 3x3x16 kernels, 32 3x3x32 kernels, and 64 3x3x32 kernels, respectively.
Each convolutional layer is followed by a $ReLU(x) = max(0, x)$ non-linear activation function, a 1-max pooling layer with $2 \times 2$ pooling region, and a batch normalization layer~\cite{ioffe2015batch}.
%The output feature map contains 1/4 features of the input feature map after 1-max pooling.
%The batch normalization helps to reduce covariance shift in a neural network.
The decoder upsamples a value in the input kernel map into a $2 \times 2$ region of that value.
It has four upsampling layers.
After each upsampling layer, the decoder uses a transposed convolutional layer, which uses 32 3x3x64 kernels, 32 3x3x32 kernels, 16 3x3x32 kernels, 3 3x3x16 kernels, respectively.

\begin{figure*}
    \centering
    \frame{\includegraphics[width=1.0\textwidth]{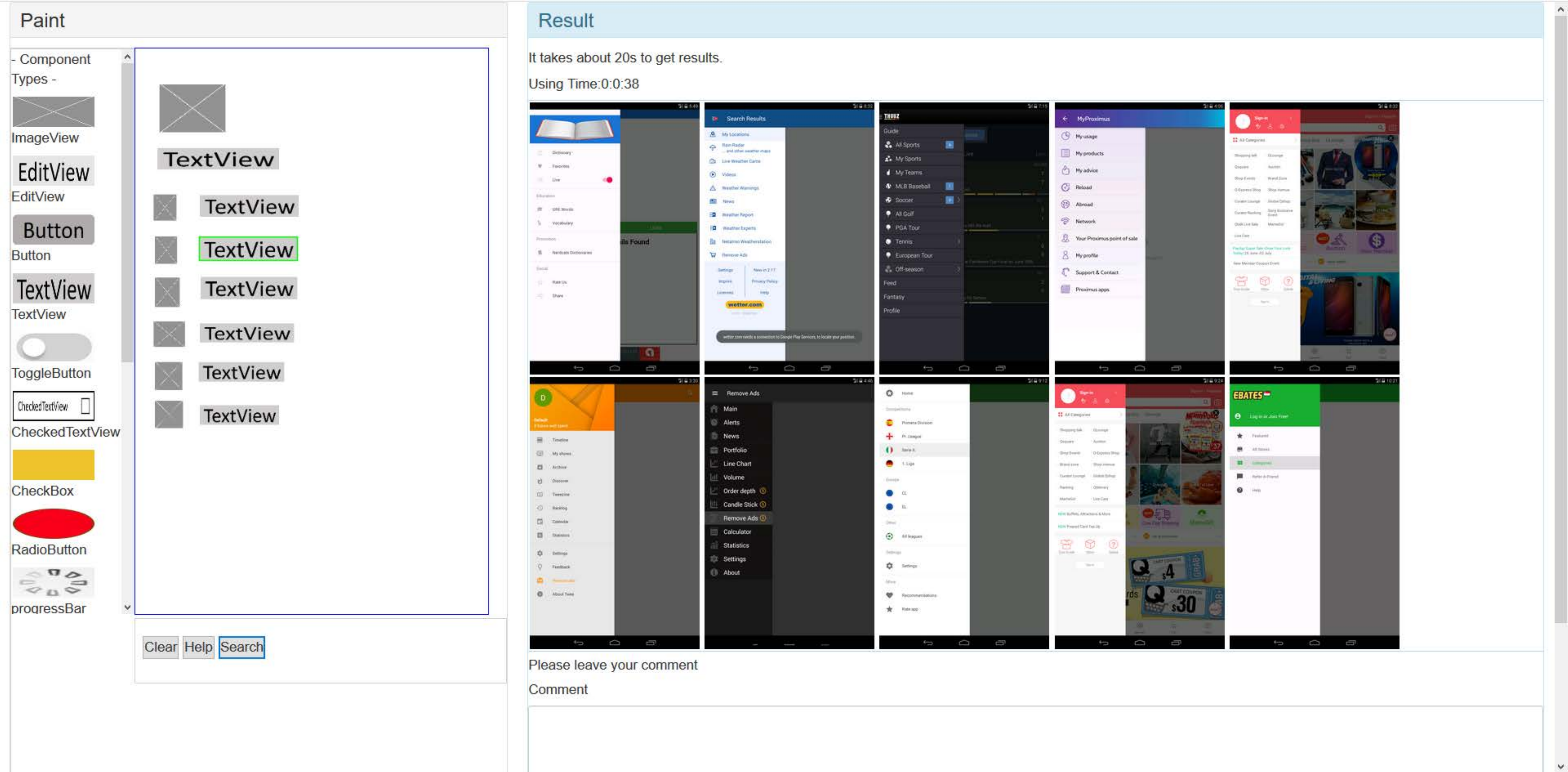}}%
    \caption{The core page of our User Interface Search Website.}
    \label{fig:website}
\end{figure*}

\subsection{Tool Implementation}
We use k=10 for KNN in all our experiments and our tool\footnote{We do not make K too big as developers tend not to browse a long list of recommendations~\cite{thung2013automated,chen2019mining, li2019linklive, xu2018domain, xu2017answerbot}.}.
Figure~\ref{fig:website} shows the frontend of our tool.
%The frontend of our tool is a web-based UI design search interface.
A demo video of this search interface is available in our Github repository\cref{web2}, which demonstrates the UI design search process of our motivating scenario.
Using our tool, the user draws a UI wireframe on the left canvas.
The tool currently supports 16 most frequently-used types of wireframe components as shown in Figure~\ref{fig:wiriframeComponents}.
We identify these wireframe components as core for Android mobile applications by surveying Android GUI framework and popular UI design tools such as Adobe XD~\cite{AdobeXd}, Fluid UI~\cite{FluidUI}, Balsamiq Mockups~\cite{BalsamiqMockups} as stated in Section~\ref{sec:wirification}.
Once the user clicks search button, the system returns the top-10 (i.e., k=10 for KNN) UI designs in the UI design space that are most similar to the wireframe on the drawing canvas.
The user can iteratively refine the wireframe and search relevant UI designs.

\section{Experiment Design}
\label{sec:experiments_setup}
In this section, we describe the research questions (RQs) we investigate in experiments and then elaborate the setup for each RQ, including the experimental dataset, baseline models, metrics and procedure.

\subsection{Research Questions (RQs)}
\label{sec:rq}
Our evaluation aims to answer the following RQs:

\begin{itemize}
 \item \textbf{RQ1: Effective of the different representation of wireframes:} Which kind of color palates used to represent the wireframe performs the best? Why does the performances differ?
 \item \textbf{RQ2: Accuracy performance on artificial dataset of our UI search engine:} How well does our approach achieve the goal of finding the relevant UI designs in face of the great variations in UI design? How well does it compare with image-similarity based, component-matching based, or naive neural network-based UI design search, and what are the reasons for this?
 \item \textbf{RQ3: Generalization ability to real world user interfaces:} How well does our model perform from the perspective of developers? How well does it compare to the best baseline in RQ2?
 \item \textbf{RQ4: User study of the recommendations from our tool in terms of usefulness and diversity:} Does our search engine really help developers to design UIs? To what extent does our search engine help them in terms of the usefulness and diversity of the recommended UIs?
\end{itemize}

\begin{figure*}
	\centering
	\subfigure[Original]{%
		\frame{\includegraphics[width=0.23\textwidth]{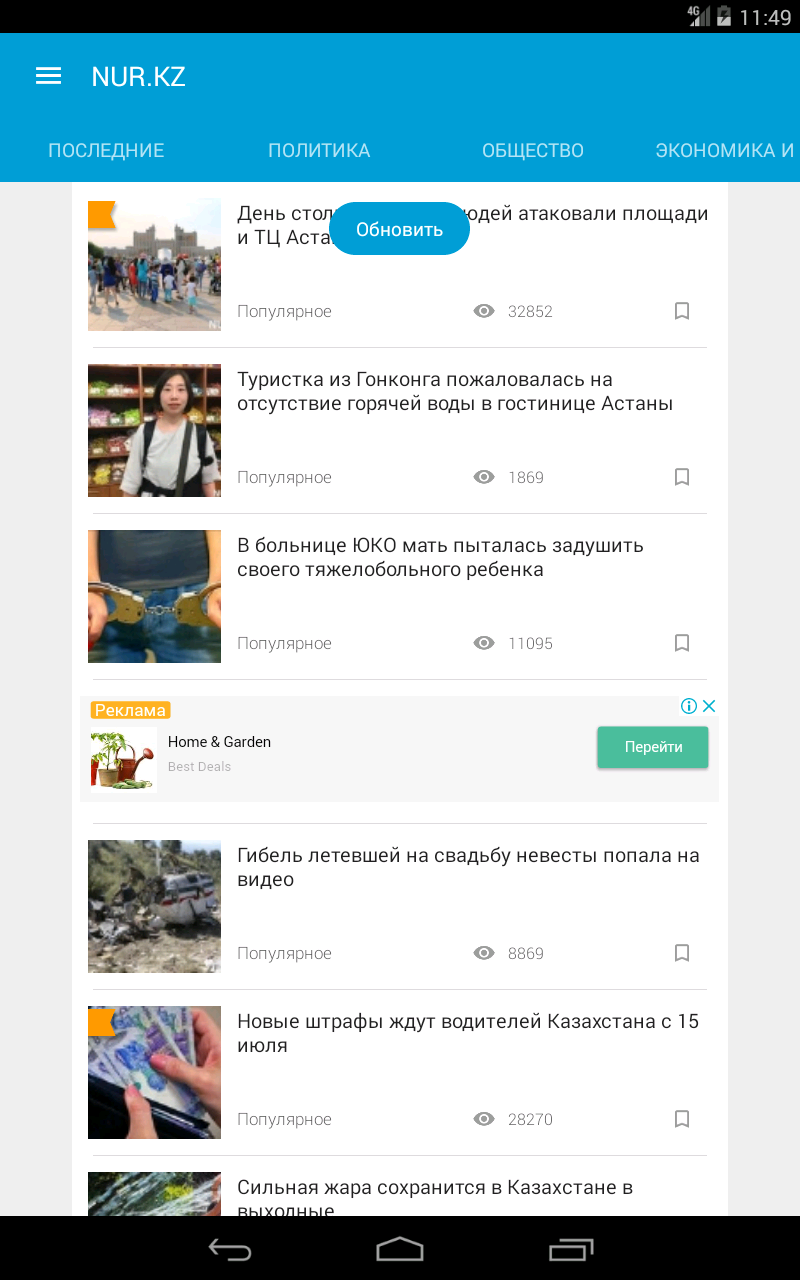}}%
		\label{fig:Original-UI}%
	}	
    \hfill
	\subfigure[Grey-level]{%
		\frame{\includegraphics[width=0.23\textwidth]{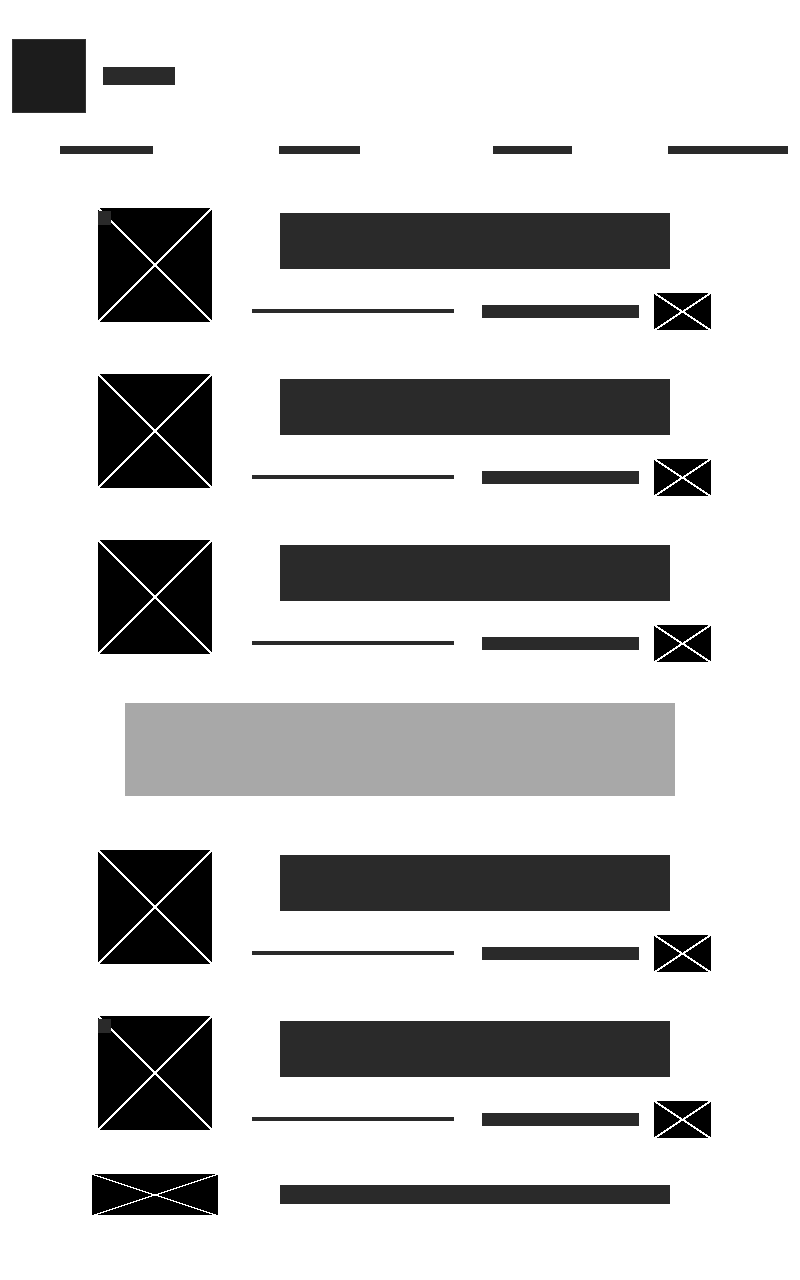}}%
		\label{fig:Grey-level}%
	}
	\hfill
	\subfigure[Color-level]{%
		\frame{\includegraphics[width=0.23\textwidth]{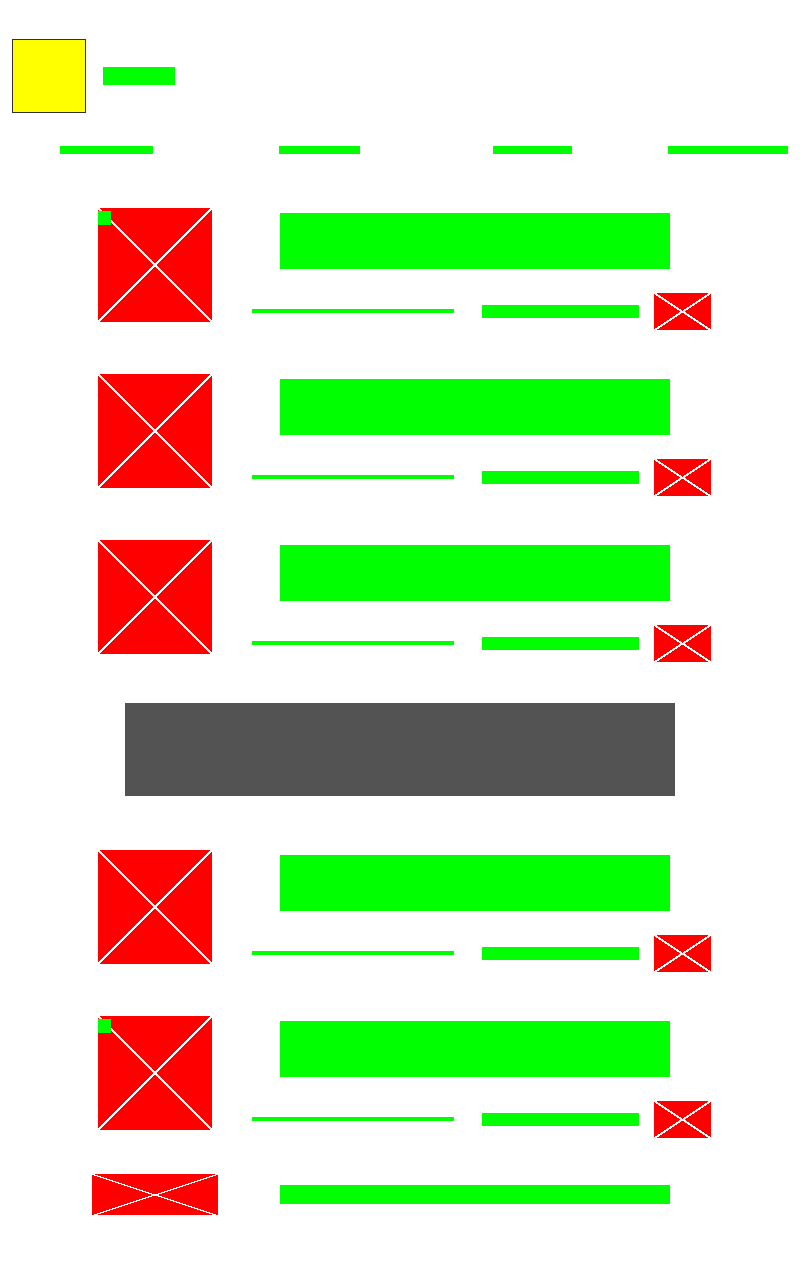}}%
		\label{fig:Color-level}%
	}
	\hfill
	\subfigure[Texture-level]{%
		\frame{\includegraphics[width=0.23\textwidth]{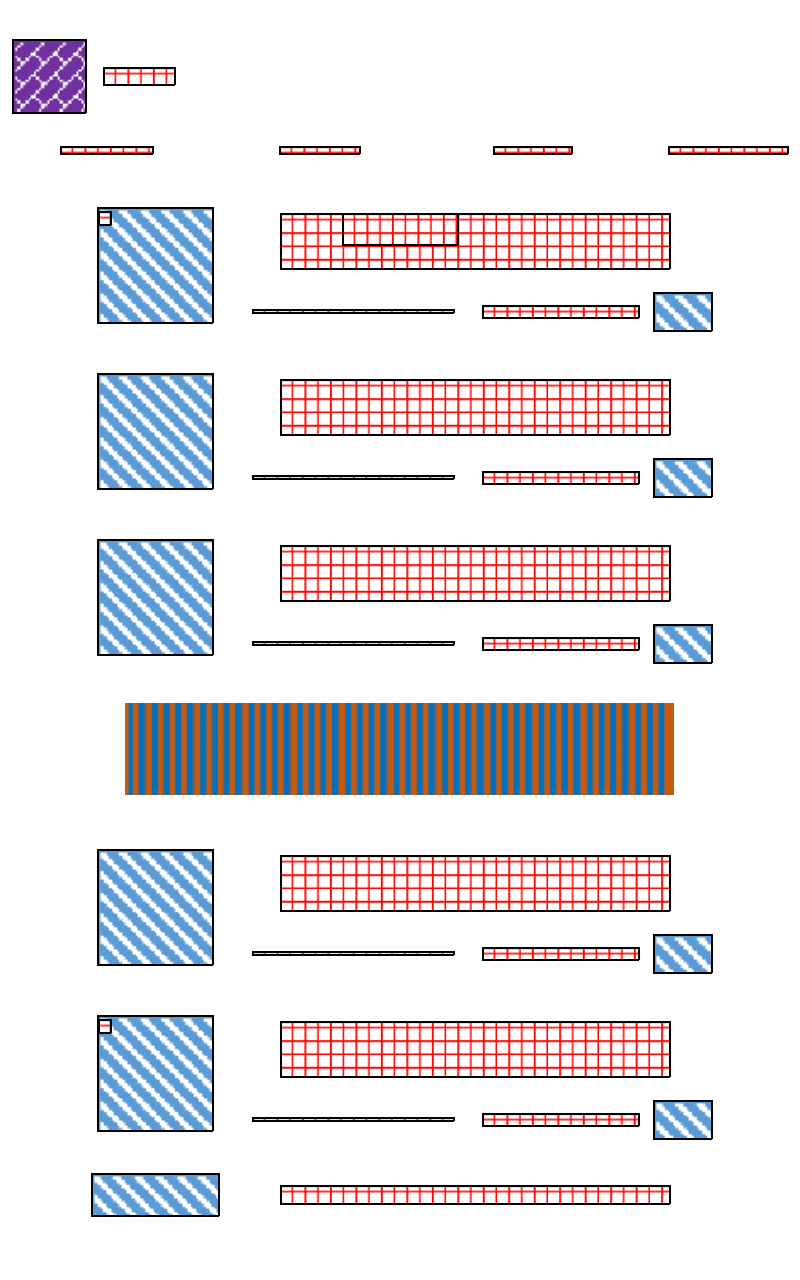}}%
		\label{fig:Texture-level}%
	}
	\caption{Example of different input format. From left to right, they are original UI images, color-level wireframe, grey-level wireframe and texture-level wireframe}
	\label{fig:ThreeRepresentation}
\end{figure*}

\subsection{RQ1: Effective of the different representation of wireframes}
\label{sec:ChoiceOfColor}
We introduce the training datasets and the experimental datasets for evaluating the effectiveness of different kinds of wireframes, and then the metrics used in this evaluation.
Note that the method used to generate these datasets and the metrics used in this evaluation are also used to answer RQ2.

\subsubsection{Experimental Dataset}
\label{sec:treatments}
To answer RQ1, we need to first construct several wireframe datasets using different representations of wireframes as the training datasets.
We then construct another experimental dataset to automatically evaluate the effectiveness of the three models trained by these different kinds of representations of wireframes.

We investigate three types of representation of visual components, including different grey-scale values, different colors and different colors with different textures.
We denote these as grey-level, color-level and texture-level wireframes respectively.
An example of these three representations can be seen in Figure~\ref{fig:ThreeRepresentation}.
For three kinds of training dataset, we can directly generate them using the method stated in Section~\ref{sec:wirification}.

\begin{figure*}
	\centering
	\subfigure[Examples of component-scaling treatment]{%
		\includegraphics[width=0.7\textwidth]{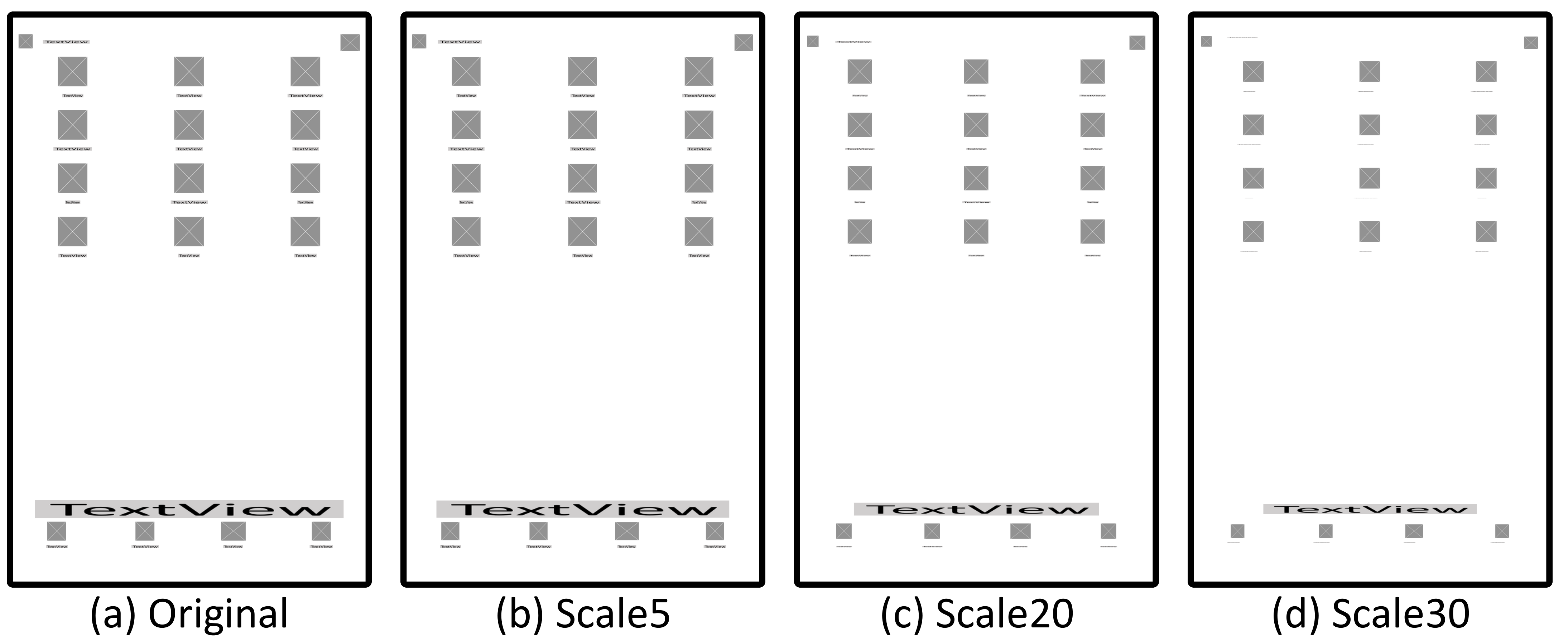}%
		\label{fig:scalingexample}%
	}
	%\hfill
	\subfigure[Examples of component-removal treatment]{%
        \centering
		\includegraphics[width=0.7\textwidth]{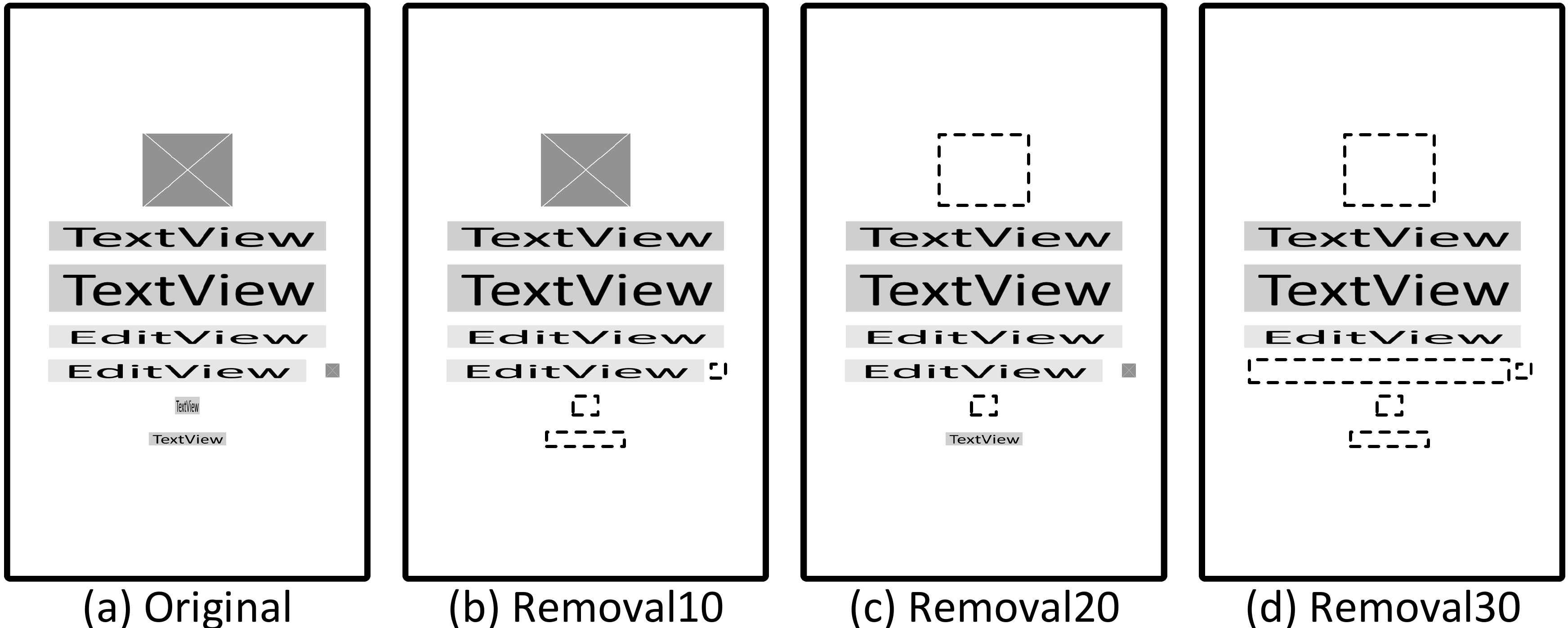}%
		\label{fig:removalexample}%
	}
	\caption{Examples of component-scaling and component-removal treatment}
	\label{fig:treatmentexample}
\end{figure*}

Second, to evaluate the performance of a UI-design search method in terms of different representations of wireframe, we require a dataset of relevant UI designs.
Unfortunately, no such datasets exist.
It is also impossible to manually annotate such a dataset in a large UI design database for large-scale experiments of a method's capability in face of different UI design variations.
Inspired by the data augmentation methods used for enhancing the training of deep learning models~\cite{perez2017effectiveness, cubuk2018autoaugment}, we change a UI screenshot in our Android UI design database to artificially create pairs of relevant but variant UI designs.

%Although there are many operations widely used in the data augmentation like resizing or rotating,
Based on the position/size of components in a UI screenshot (see Section~\ref{sec:uidesigndatabase}), we perform two types of change operations which are suitable for UI designs: \textit{component scaling} and \textit{component removal}.
Component scaling is to scale down all visual components in a UI wireframe to their center point by 5\%, 10\%, 15\%, 20\%, 25\%, or 30\% pixels (round-up) of their original size (see Figure~\ref{fig:scalingexample}) which simulates design variations in component position/size.
Component removal is to randomly remove some visual components that cover 10\%$\pm5$ \%, 20\%$\pm5$\% or 30\%$\pm5$\% of the total area of all components (see Figure~\ref{fig:removalexample}) which simulates design variations in component type/number.
We denote 10\%$\pm5$ \%, 20\%$\pm5$\% or 30\%$\pm5$\% in component removal treatment as removal10, removal20 and removal30 for simplicity.
As the examples in Figure~\ref{fig:storyLine} show, such design variations are commonly present in relevant UI designs.
In reality, these two types of design variations may occur at the same time.
But we perform the two types of changes separately to investigate a search method's capability of handling different types of design variations.

We randomly select two sets of screenshots from 25 categories, and each set is comprised of 500 screenshots.
Note that for each set, the proportion of the screenshots taken from each category is the same as the original proportion of each category in the total database.
We then apply the six scaling treatments to the first set and the three component-removal treatments to the second set.
The UI screenshots in the second set should have at least 5 UI components so that there are some components left after component removal.
As a result, we obtain 4500 pairs of original-treated UIs, which are considered as relevant but variant UI designs.
We have nine experiments (one for each treatment).
For example, the Scale10 experiment uses the UIs obtained by 10\% component scaling as query.
Note that we generate the corresponding experimental wireframe dataset three times using the above mentioned three types of representation of wireframes, which means that we have three experimental datasets, each of them contains 4500 pairs of original-treated UIs.
Using this dataset, we evaluate how well a search method can retrieve the original UI in the database using a treated UI design as query in terms of different representation of wireframes.

\subsubsection{Evaluation Metrics}
\label{sec:metrics}

We evaluate the performance of a UI-design search method by two metrics:
Precision@k (Pre@k) (k=1) and Mean Reciprocal Rank (MRR).
The higher value a metric is, the better a search method performs.
Precision@k is the proportion of the top-k results for a query UI that are relevant UI designs.
As we consider the original UI as the only relevant UI for a treated UI in this study, we use the strictest metric Pre@1: Pre@1=1 if the first returned UI is the original UI, otherwise Pre@1=0.
MRR computes the mean of the reciprocal rank (i.e., 1/r) of the first relevant UI design in the search results over all query UIs.

\subsection{RQ2: Accuracy performance of our UI search engine}
In this RQ, we use the explored best representation of wireframes in RQ1.
We evaluate  how well our approach achieves the goal of finding the relevant UI designs in the face of great variations in UIs and how well it compares with image-similarity based, component-matching based, or naive neural network based UI design search, and what are the reasons behind this.
In the following, we introduce the experimental dataset and metrics used in this RQ and then elaborate the baseline models used.

\subsubsection{Dataset and Metrics}
To automatically evaluate and compare the effectiveness of our model and the baselines, we use the same treatments to construct the experimental dataset, while at this time, we only need to consider one kind of representation of wireframes.
We again randomly select 500 UI images from 25 categories, which are different from the data in RQ1, to construct the experimental dataset, and then apply the nine treatments stated in Section~\ref{sec:treatments}.
In total, we have 4500 pairs of original-treated UIs, which are considered as relevant but variant UI designs.
Using this dataset, we evaluate how well a search method can retrieve the original UI in the database using a treated UI design as query.
Beside, we also take the same metrics stated in Section~\ref{sec:metrics}

\subsubsection{Baseline Methods}
\label{sec:baselines}
We consider four baselines: two of them computes image similarity using simple color histogram and advanced SIFT feature respectively, the third one implements the component-matching heuristics proposed by GUIFetch~\cite{behrang2018guifetch}, and the last one uses the naive neural network with fully connected layers from Rico~\cite{deka2017rico}.
These four baselines are used to evaluate RQ2, and the results of RQ2 would determine the baselines used in RQ3.

\textbf{Image-feature based similarity.}
Color histogram is a simple image feature that represents the distribution of colors in each RGB (red, green, blue) channel.
It has been widely used for image indexing and image retrieval~\cite{jain1996image, han2002fuzzy, hafner1995efficient}.
The scale-invariant feature transform (SIFT)~\cite{lowe1999object} is an advanced image feature widely used for image retrieval~\cite{ke2004efficient, wu2009bundling}, object recognition~\cite{lowe1999object}, image stitching~\cite{brown2007automatic}.
It locates keypoints in images and use the local features of the keypoints to represent images.
Different images can have different numbers of keypoints but each keypoint is represented in a same-dimensional feature vector.
These two baselines return the top-k most similar UI designs by the image-feature similarity.

\textbf{Heuristic-based component matching.}
GUIFetch~\cite{behrang2018guifetch} is a recently proposed technique for searching similar GUIs by a similarity metric computed from the matched components between the two GUIs.
It matches the components of the same type.
The similarity of the two components is calculated based on the differences of the two components' x-coordinate, y-coordinate, length and width.
If the difference of one factor is within a given threshold, the similarity score increases by 10, otherwise 0.
After computing the similarity score for each pair of components in the two UIs, it uses a bipartite matching algorithm~\cite{karp1990optimal} to determine an optimal component matching.
The similarity scores of the matched components are summed up and then divided by the maximum similarity value that the components in the query UI can have (i.e., 40 multiplies the number of components in the query UI).
It then returns the top-k UIs with the highest similarity scores to the query UI.

\textbf{Neural-network-based matching.}
Rico~\cite{deka2017rico} is a UI dataset introduced to support various tasks in the UI design domain.
It demonstrates the potential usage of UI search based on a naive neural network with six fully connected layers within the autoencoder framework.
The latent vectors from their model are used as the features of their wireframe dataset.
In terms of inference, they first extract the latent vector of the query wireframe, compare it with the latent vectors of their dataset, and then return the nearest neighbors as recommendations.
For a fair comparison, we adopt the same configuration mentioned in their paper for training the model on our dataset."

\subsection{RQ3:Generalization ability to real world user interfaces}
To further evaluate our model when applied to real world applications, we conduct a human evaluation of the relevance of the UI design search results.
We do this by searching the UI design database using unseen UI designs as queries to confirm the generalization of our model and answer RQ3.
To this end, we need to construct another dataset of unseen UIs.
We introduce this dataset, the human evaluation procedure and the metrics in the following.
Note that the baselines in this RQ are defined by the results of RQ2.
We only consider several (not all) baselines from RQ2, because human evaluation of the relevance of UI design search results is labor intensive.

\subsubsection{Dataset of Unseen Query UIs}
The UI design database of our tool contains UI screenshots from 25 categories of Android applications.
We randomly download one more application per category which have not been included in our proof-of-concept implementation.
The same reverse-engineering method~\cite{chen2018ui} is used to obtain the UI screenshots of this newly downloaded application.
We generate the corresponding UI wireframes for the collected UI screenshots as described in Section~\ref{sec:uidesigndatabase}.
We select two UI wireframes per application and obtain 50 UI wireframes as the query UI design in this study.
The selected UI wireframes contain variant types and numbers of visual components, according to our observation.

\subsubsection{Human Evaluation of UI Design Relevance}
We recruited five participants, P1, P2, P3, P4 and P5, from our school as human annotators.
They have been working on Android app development for at least two years.
For each query UI, we obtain the top-10 search results (i.e., 20 UI designs in total) by our method and the GUIFetch baseline respectively.
To avoid expectancy bias, these 20 UI designs are randomly mixed together so that the human annotators have no knowledge about which UI design is returned by which method and the ranking of that UI design in the search results.
The two annotators examine the UI design search results independently.
For each query UI wireframe, they classify each of the 20 UI designs as relevant or irrelevant to the query UI.
The annotators are given the original UI screenshot of the query UI wireframe as a reference for comparison.

\subsubsection{Evaluation Metrics}
\label{sec:rq3_metrics}
We use two statistical methods to measure the inter-rater agreement between two human annotators and among all five human annotators.
For the first metric, we compute Cohen's kappa statistics~\cite{cohen1960coefficient}, which is suitable for measuring the agreement between two raters accessing multiple items into two categories.
For the second metric, we compute Fleiss's kappa statistics~\cite{fleiss1971measuring}, which is used to evaluate the agreement between multiple raters.
Based on the five annotators' judgment of UI design relevance,
we regard a returned UI design as relevant by three strategies: strict (both annotators label it as relevant) moderate (the majority of annotators label it as relevant) and relaxed (at least one annotator labels it as relevant).
We then compute Precision@k (k=1, 5, 10) and MRR.
We do not use Recall and Mean Average Precision (MAP) in this study because it is impossible to manually annotate all relevant UI designs for a query UI in a large UI-design database (54,987 UI screenshots in our proof-of-concept implementation).

\subsection{RQ4: User study of the recommendations from our tool in terms of usefulness and diversity}
To answer RQ4 and evaluate the usefulness of our search engine, we conduct a user study.
We choose five UI design tasks from Daily UI design challenge\footnote{\label{dailyui}\textcolor{blue}{\url{https://www.dailyui.co/}}}, recruit 18 students to design these tasks, search relative UIs and modify their draft using our tool, and then rate the usefulness and diversity of the recommendations.
In the following, we introduce the details of these five tasks, the experiment procedure and the metrics used in this RQ.

\subsubsection{UI Design Tasks}
We select five UI design tasks from Daily UI design challenge\cref{dailyui}: \textit{sign-up}, \textit{image gallery}, \textit{login}, \textit{preference setting}, and \textit{navigation drawer}.
These five UIs cover essential features of Android mobile applications:
\textit{sign-up} and \textit{image gallery} are typical UIs for collecting user inputs and displaying information content, respectively.
\textit{login} is a common feature for user authentication, and \textit{preference setting} is commonly used for software customization.
\textit{navigation drawer} is a core interaction feature to provide users the access to all app functionalities.
Furthermore, these features are easy to understand even for non-professional UI designers who are the targeted users in this study

\subsubsection{Experiment Procedure}
We recruit 18 students from our school through the school's mailing list.
Although six students have some front-end software development experience, none of the participants have Android UI design experience.
In other word, they are inexperienced designers, same as Lucy in Section~\ref{sec:motivation}.
Participants are given the five UI design tasks and are asked to design a UI wireframe for each task.
Each task is allocated 15-30 minutes.
Due to the time limitation, we do not ask the participants to design high-fidelity visual effects of the UIs.
To assist their design work, the participants use our web tool to draw the UI wireframes and search our database of 54,987 Android UI designs (see Section~\ref{sec:Implementation}).
The tool returns the top-10 UI designs for a query UI.
We give the participant a tutorial of tool usage and a 15-minute warm-up session to learn to use the tool.
For each task, the participants can search as many times as they wish.
For the last search, they are asked to select the UI designs in the search results that they consider relevant to the query UI wireframe they draw.
They are also asked to rate the overall diversity and usefulness of the search results of the last search by 5-point Likert scale (1 being the lowest and 5 being the highest).
In detail, usefulness refers to how useful search results help participants understand/adjust design options if they are facing real UI design tasks.
For example, when participants are searching for some UIs related building a sign-up page, the recommendations from our search engine fit into their design requirements.
Diversity refers to the diversity of the recommendation results, for example, whether the recommended UIs involve variant component usage/layouts or color/size/font effects which may be beyond their expectations.

\subsubsection{Evaluation Metrics}
We record the times of search by the participants for each task.
Based on the relevance judgment of UI design search results for the last search, we compute Precision@k (k=1, 5, 10) and MRR.
We do not report Recall and MAP as it is impossible to annotate all relevant UI designs in our database of 54,987 UI screenshots for a user-drawn UI wireframe.

\begin{figure*}
	\centering
	\includegraphics[width=1.0\textwidth]{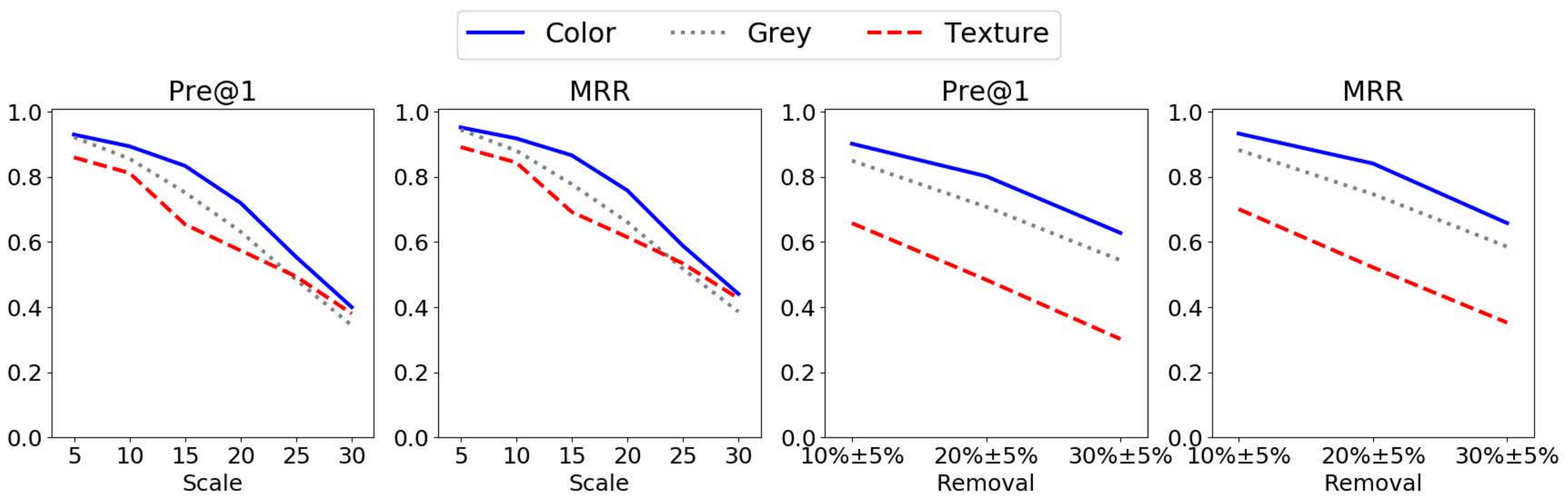}
	\caption{Results of three types of representation}
	\label{fig:different_representation_results}
\end{figure*}

\begin{figure*}
    \centering
    \includegraphics[width=0.95\textwidth]{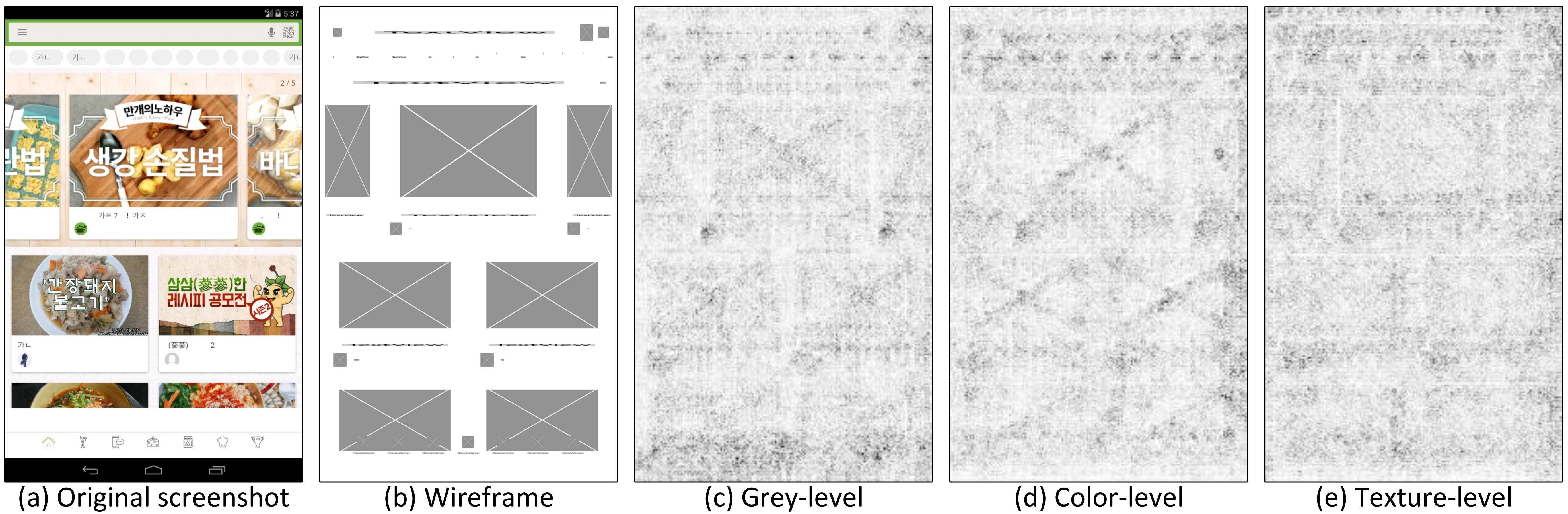}%
    \caption{Heatmaps for the grey-level, color-level and texture-level wireframe}
    \label{fig:heatmap}
\end{figure*}

\section{Experimental Results}
\label{sec:results}
\subsection{RQ1 Results}
\label{sec:rq1_results}
\subsubsection{Quantitative Results.}
Figure~\ref{fig:different_representation_results} shows the results of evaluating the three different representation wireframes. Overall, the color-level model remains a slight advantage over grey-level model in both component-scaling and component-removal treatments, with a 5\%-10\% and 0.01-0.1 increase in Pre@1 and MRR respectively.
The reason may be that the color-level model mainly focuses on the boundary of contained components in a wireframe instead of the exact pixel values, and the color-level wireframe input includes three channels encoding more information while the grey-level wireframes includes only one channel.
In contrast, there are large margins between the performances of the color-level model and of the texture-level model, especially in component-removal treatments.
This may be because the texture information is too complex and may confuse the model after several max-pooling layers.

\subsubsection{CNN Visualization.}
To better understand the impacts of different representations, we visualize these models using vanilla (i.e., standard) backpropagation saliency~\cite{chen2018data} in Figure~\ref{fig:heatmap}.
We can find that the heatmap of the color-level model is the clearest, while that of the texture-level model is the vaguest with much noise.
The grey-level heatmap is vaguer than color-level one because the differences between component and background is small in the grey one.
In conclusion, the color-level model performs the best and we choose it as the representation of our wireframe dataset.

\subsection{RQ2 Result}
\label{sec:rq2}
Our deep-learning based approach for UI design search is the first technique of its kind.
It is designed to find relevant UI designs in face of the great variations in UI designs.
In this RQ2, our goal is to evaluate how well our approach achieves this design goal, and how well it compares with image-similarity based or component-matching based UI design search.
We use the color-level wireframe dataset in this evaluation as the effectiveness of this kind of representation have been proved in Section~\ref{sec:rq1_results}.

\subsubsection{Runtime Performance}
\label{section:Overall performance}
We run the experiments on a machine with Intel i7-7800X CPU, 64G RAM and NVIDIA GeForce GTX 1080 Ti GPU.
Take the inference time of the Scale10 experiment as an example.
Our W-AE (short for Wireframe Autoencoder), Rico, GUIFetch, SIFT and color-histogram take 561.2 seconds, 771.4 seconds, 7446.9 seconds, 3944.6 seconds and 523.6 seconds for 500 queries, respectively.
In general, W-AE is about 12 times and six times faster than the GUIFetch and SIFT baselines, and is as fast as the color-histogram and Rico baselines.

\subsubsection{Retrieval Performance}
\label{sec:autoretrievalperformance}

Figure~\ref{fig:autoresults} shows the performance metrics of the five methods in the nine treated-UI-as-query experiments.
The color-histogram baseline and the SIFT baseline have close performance in all component-scaling experiments.
At the component scaling ratio 10\%, their performance metrics become lower than 0.2, and at the ratio 20\% or higher, their performance metrics become close to 0.
For component removal experiments, the advanced SIFT feature performs better than the simple color histogram feature.
This is because the UIs treated by component removal still have many intact components (see Figure~\ref{fig:removalexample}), which have the same keypoints (and thus the same SIFT features) as their counterparts in the original UIs.
This helps to retrieve the original UIs for the component-removal-treated UIs.
Nevertheless, at the component-removal ratio 20\% or higher, the performance metrics of the SIFT baseline become lower than 0.5.

In contrast, our W-AE is much more robust in face of large component-scaling and component-removal variations, because our CNN model can extract more abstract, sophisticated UI-design related image features through deep neural network, which are much less sensitive to image differences than low-level image features like color histogram or SIFT.
At the component scaling ratio 20\%, our W-AE still achieves 70.0\% Precision@1 and 0.73 MRR.
The performance of our W-AE degrades (but is still much better than the four baselines) when the component-scaling ratio is 25\% or higher.
This is because many small-size visual components (such as checkbox, switch or small text) will become hardly visible even for human eyes (see Figure~\ref{fig:scalingexample}).
Similarly, the features of such extremely-small components will become invisible to the ``eye'' (i.e., convolutional kernels) of the CNN model, and thus cannot contribute to the measurement of the UI-design similarity.
Although such extremely-small UI components can test the limits of a search method in extreme conditions, they would rarely exist in real-word UIs because they are not user friendly.
At the component-removal ratio 20\%, our W-AE still achieves 84.6\% Precision@1 and 0.88 MRR.
The model performance degrades (again still much better than the baselines) at the component-removal ratio 30\%.
However, as the example in Figure~\ref{fig:removalexample} shows, the treated UI with components covering 30\% less area than the original UI may become not-so-similar anymore to the original UI.
But we still consider the original UI as the ground truth for the treated UI in our automatic experiments, which may result in the biased metrics for all the evaluated methods.

\begin{figure*}
	\centering
	\includegraphics[width=1.0\textwidth]{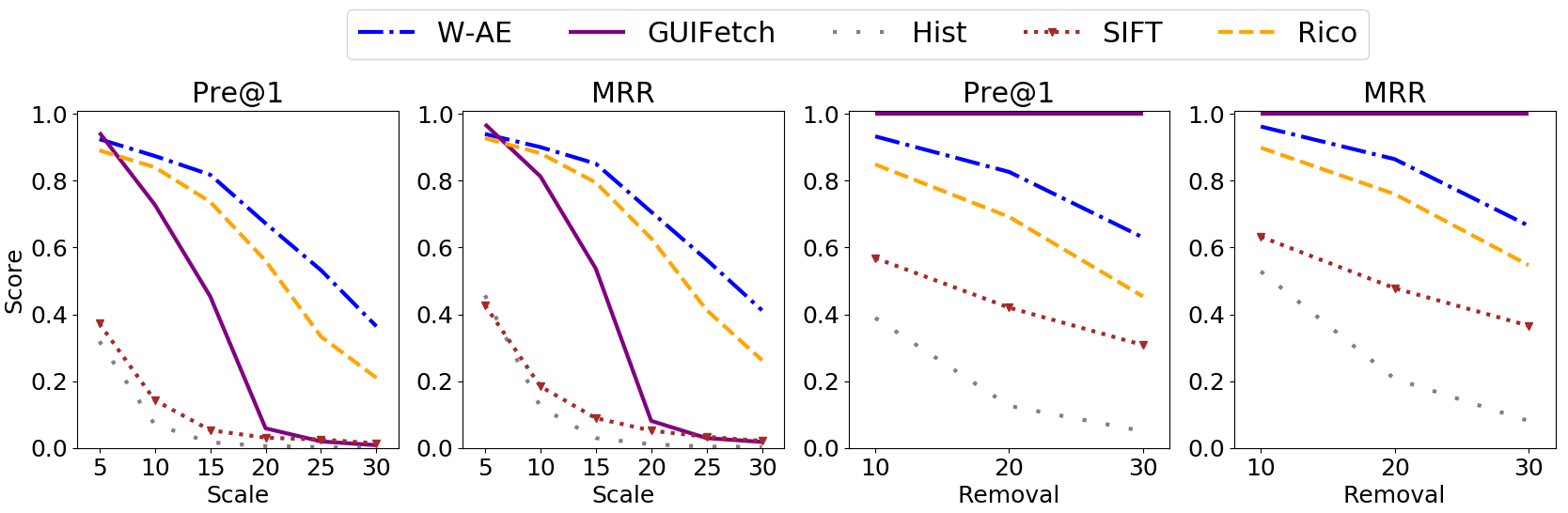}
	\caption{Results of nine automatic experiments}
	\label{fig:autoresults}
\end{figure*}

\begin{figure*}
	\centering
	\subfigure[Several well-aligned, close-by, same-type components as one component]{%
		\includegraphics[width=0.51\textwidth]{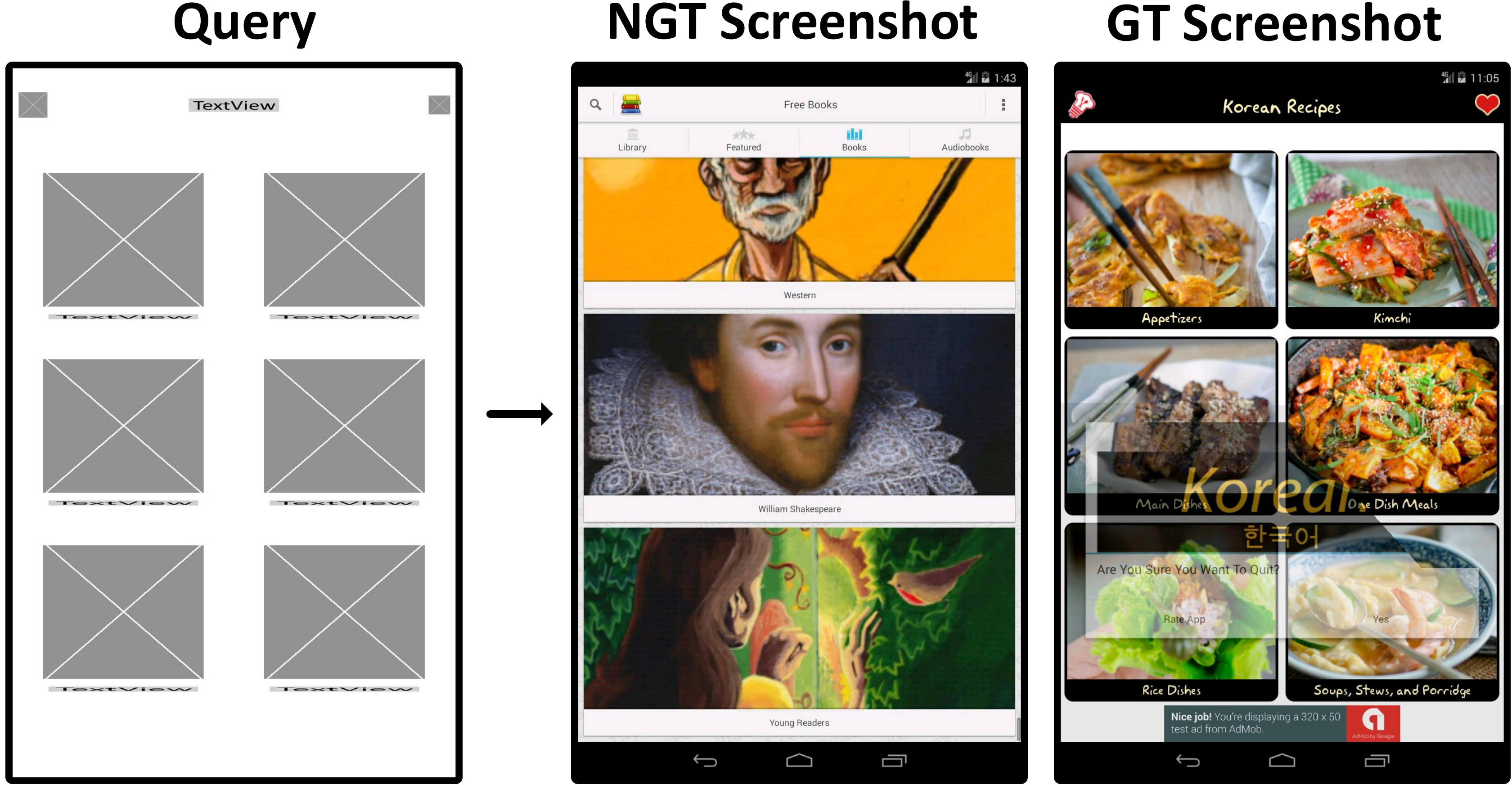}%
		\label{fig:combined widget}%
	}
	\hfill
	\subfigure[Similarity of large components overshadows that of small components]{%
		\includegraphics[width=0.51\textwidth]{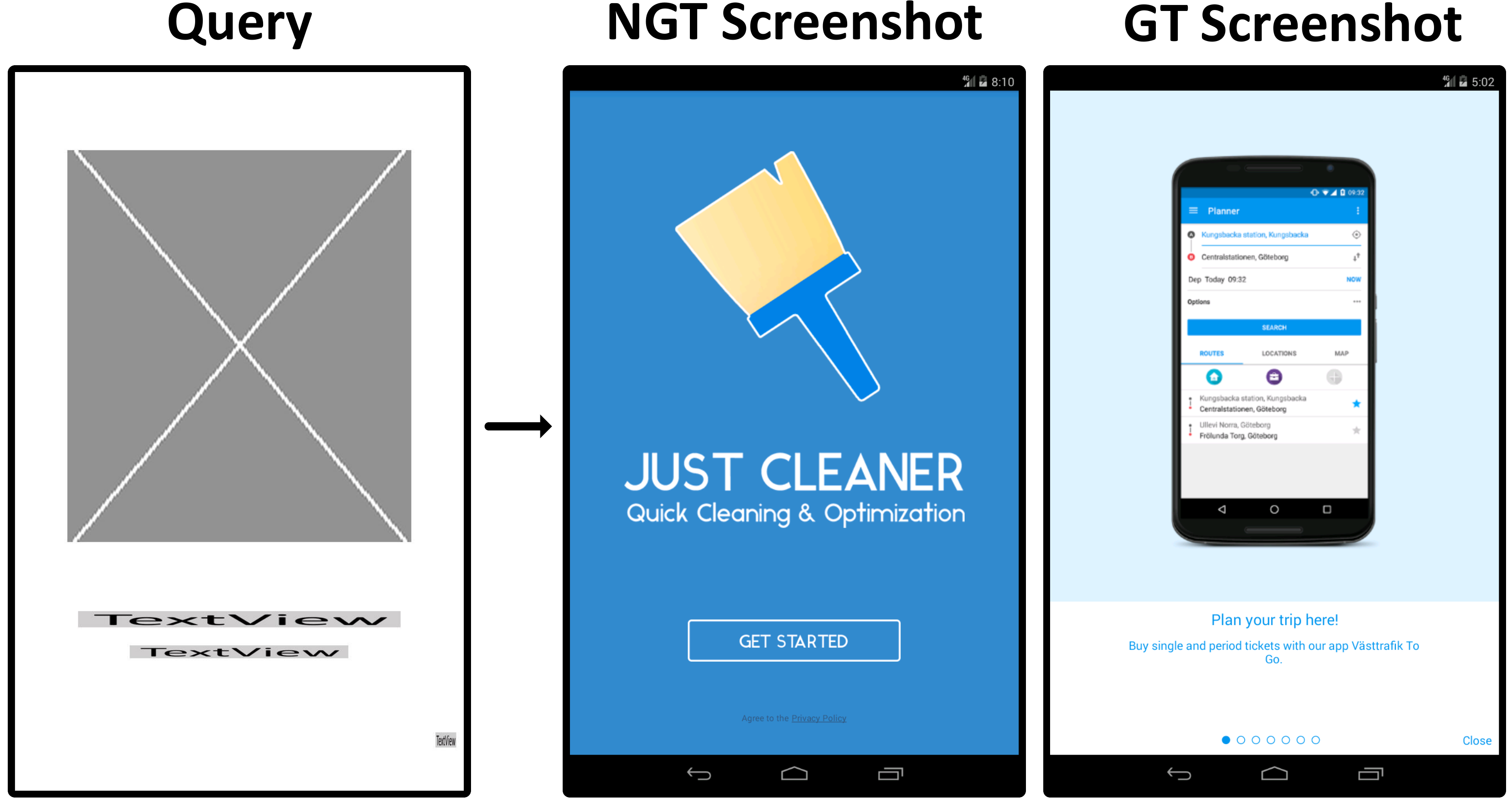}%
		\label{fig:big widget accounts for more loss}%
	}
	\hfill
	\subfigure[Foreground components are overlooked]{%
		\includegraphics[width=0.51\textwidth]{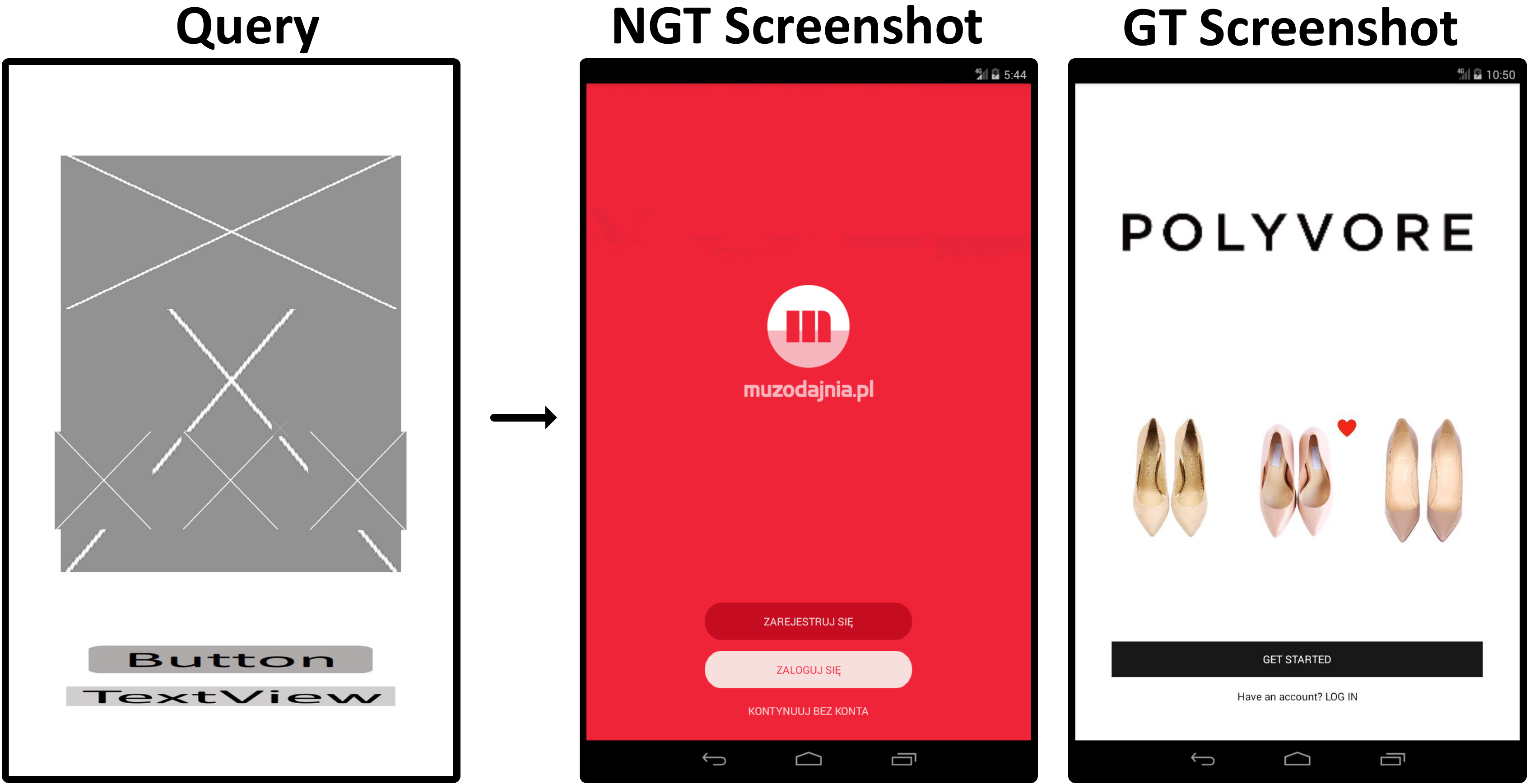}%
		\label{fig:same color widgets overlay}%
	}
	\caption{Examples of non-ground-truth (NGT) UI ranked before ground-truth (GT) UI}
	\label{fig:Common causes}
\end{figure*}

Our W-AE outperforms Rico on all metrics by large margins.
The Rico baseline performs better than other baselines but the performance gap between Rico and our W-AE model keeps growing as the UI components and layout similarity decreases.
Within the component-scaling experiments, the Rico baseline is comparable to our W-AE at the scaling ratio 5\%-10\%, but degrades quickly when the ratio is 25\% or higher.
This is because Rico applies fully connected layers, which consider every pixel in the UIs without filtering out meaningless and noisy ones, leading to sensitivity to small input changes.
In comparison, our W-AE performs convolution and pooling strategies to extract the core features from UIs which is much more stable.
A similar observation also applies to experiments of removal treatment.
The fully connected neural network baseline achieves relatively good performance at the ratio 10\% and then drops to 70.6\% Precision@1 and 0.77 MRR at the ratio 20\%, and 47.6\% Precision@1 and 0.57 MRR at the ratio 30\%.

Among all component-scaling experiments, the GUIFetch baseline achieves comparable performance as our W-AE only at the scaling ratio 5\%.
However, the performance of the GUIFetch baseline drops significantly when the component-scaling ratio increases, and becomes close to 0 at the scaling ratio 20\% or higher.
This is because of the sensitivity of the GUIFetch's component matching rules (see Section~\ref{sec:baselines}).
When the component-scaling ratio is large, the position and size of the corresponding components in the treated UI and the original UI will no longer be close enough under the threshold, and thus will not be matched.
For all component-removal experiments, the GUIFetch baseline ``unsurprisingly'' achieves the perfect performance (all metrics being 1.0).
This perfect performance is because all components left in a treated UI are intact and thus can match their counterpart components in the original UI.
Furthermore, the GUIFetch's similarity metric considers only the matched versus unmatched components in the query UI.
As such, the treated UI and the original UI end up with the similarity score 1.0.
However, our generalization study shows that the component-matching heuristics and the similarity metric of GUIFetch do not work well in reality for finding relevant UI designs for real-world query UIs as judged by human.

\subsubsection{Retrieval Failure Analysis}
\label{section:common reason}
To gain deeper insight into our CNN model's capability of encoding the visual semantics of UI designs, we manually examine the retrieval-failure cases in which the ground-truth (GT) UI is ranked after other non-ground-truth (NGT) UIs, and identify three main causes for retrieval failures in our automatic experiments.
Figure~\ref{fig:Common causes} shows the typical examples for these three types of retrieval failures.

First, the query UI contains several well-aligned, close-by, same-type components, but the model returns some UIs that have some same-type but bigger and less number of components in the corresponding UI region (Figure~\ref{fig:Common causes} (a)).
This reveals the limitation of our model in distinguishing several well-aligned, close-by, same-type components from one another.
However, certain level of modeling fuzziness is important for retrieving similar UI designs with variant numbers of components (such as Figure~\ref{fig:storyLine} (g) versus (h)/(i)/(j) and the Image Gallery example in Figure~\ref{fig:results of userstudy2}).
It is important to note that we use pairs of the original and treated UIs as relevant UI designs in our automatic experiments, and consider all other UIs in the database as ``irrelevant'' for a query UI.
However, as the example in Figure~\ref{fig:Common causes} (a) shows, the non-ground-truth UIs can still relevant to the query UIs.
Such UI design relevance can only be judged by human, as we do in the generalization experiment and user study.

Second, the query UI contains a UI component (usually an ImageView) covering a large area of the UI, and the model returns some UIs that are similar to the query UI only by that large component, but not similar in other parts of the UI designs (Figure~\ref{fig:Common causes} (b)).
Such retrieval results indicate that our model does not treat the features from small or large visual components equivalently.
This inequivalent treatment is reasonable as large components are visually more evident, but it may result in the similarity of large components overshadowing that of small components.

Third, the query UI contains some foreground UI components overlapping a large background component, but the returned UIs contain only the background component without the foreground components, especially when the foreground and background components are of the same type.
Overlapping components, especially the same-type ones, can be visually indistinguishable in UI wireframes, because they lack high-fidelity visual effects (e.g., distinct colors or images) to tell them apart.
They pose a threat to our wireframe-based UI design search.
However, according to our observation, most of UI designs with overlapping components have a background image on top of which real-functional UI components are laid.
By removing such background images when generating the UI wireframes, this threat could be mitigated.

\vspace{2mm}
\noindent\fbox{\begin{minipage}{13.5cm} \emph{In the face of component-scaling and component-removal variations in UI designs, our CNN-based method that models the visual semantics of the whole UI designs significantly outperforms the image-similarity based and the component-matching based methods.
But the performance of our method could be further enhanced by the capability of modeling well-aligned, close-by components, small-size components, and overlapping components.} \end{minipage}}\\

\subsection{RQ3 Results: Generalization Evaluation}
\label{sec:Generalization}

Based on the performance results of the three baseline methods in our automatic evaluation in Section~\ref{sec:rq2}, we use the GUIFetch baseline in this study.
We do not use the color-histogram and SIFT baselines for two reasons.
First, our automatic evaluation shows that the color-histogram and SIFT baselines have very poor performance even in face of artificial design variations.
Second, human evaluation of the relevance of UI design search results is labor intensive, and considering two more baselines will double the manual evaluation effort.

\begin{table*}
	\centering
	\caption{Pairwise comparisons of inter-rater agreements}
    %\footnotesize
    \setlength{\tabcolsep}{1.2em}
    \renewcommand{\arraystretch}{1.2}
	\begin{tabular}{|l| c| c| c| c |c |}
        \hline
                    & \textbf{P1} & \textbf{P2} & \textbf{P3} & \textbf{P4} & \textbf{P5}  \\
        \hline
        \textbf{P1} & -   & 0.43 & 0.37 & 0.45 & 0.48 \\
        \hline
        \textbf{P2} & 0.43 & -    & 0.38 & 0.51 & 0.49\\
        \hline
        \textbf{P3} & 0.37 & 0.38 & -    & 0.43 & 0.42 \\
        \hline
        \textbf{P4} & 0.45 & 0.51 & 0.43 & -    & 0.56 \\
        \hline
        \textbf{P5} & 0.48 & 0.49 & 0.42 & 0.56 & - \\
		\hline
	\end{tabular}
	\label{tab:kappa}
\end{table*}

\begin{table}
	\centering
	\caption{Results of human relevance evaluation }
    %\footnotesize
	\begin{tabular}{|l|cc|cc|cc|cc|}
        \hline
        \multirow{2}{*}{} &
            \multicolumn{2}{c|}{\textbf{Relaxed}} &
            \multicolumn{2}{c|}{\textbf{Moderate}} &
            \multicolumn{2}{c|}{\textbf{Strict}} \\
            & \textbf{W-AE} & \textbf{GUIFetch} & \textbf{W-AE} & \textbf{GUIFetch} & \textbf{W-AE} & \textbf{GUIFetch}  \\
        \hline
        \textbf{Pre@1}  & \textbf{0.84} & 0.64   & \textbf{0.5}  & 0.32 & 0.14          & \textbf{0.16} \\
        \textbf{Pre@5}  & \textbf{0.77} & 0.65   & \textbf{0.47} & 0.34 & \textbf{0.20} & 0.13 \\
        \textbf{Pre@10} & \textbf{0.75} & 0.62   & \textbf{0.43} & 0.31 & \textbf{0.15} & 0.12 \\
        \hline
        \textbf{MRR}    & \textbf{0.90} & 0.78   & \textbf{0.62} & 0.48 &\textbf{0.27}  & 0.24 \\
		\hline
	\end{tabular}
	\label{tab:Generalization Results}
\end{table}

\subsubsection{Results}

All participants spent about 120 minutes to rate the relevance of the 1000 UI designs to their corresponding query UI wireframes.
Table~\ref{tab:kappa} shows the Cohen's kappa results of the pairwise comparisons among all five participants.
For these comparisons, most of the kappa statistics fall in the range of 0.42-0.56, which indicates a moderate to substantial agreement.
We further conducted the Fleiss's kappa~\cite{fleiss1971measuring} to evaluate the agreement among all raters.
The Fleiss's kappa for the 2500 (500x5) annotations of UI design search results by our method and the GUIFetch baseline are 0.41 and 0.51, respectively.
We consider this level of agreement as acceptable because it can be rather subjective for determining the relevance of UI designs, depending on different background, experience, education and even culture of the human annotators.

According to our observations and interviews, there are four aspects these annotators that are most concerned with, including the semantic meaning of the UI (i.e., functionality), the layout, the types of components and the number of components.
Some participants focus more on some aspects while others are more concerned with other parts.
Some participants are rather strict while others are relatively relaxed.
Different from manual-labelling tasks like image classification, there is no hard right or wrong answer for checking each recommendation result.
Therefore, the Cohen's and Fleiss's kappa rates are not so high.
However, in summary, the overall feedback quantitative results from all participants still reflect that our method is much better than the baseline as we discuss later.

Table~\ref{tab:Generalization Results} shows the performance metrics of our method and the GUIFetch baseline. Our W-AE significantly outperforms the GUIFetch baseline in relaxed and moderate strategies by a large margin, and maintains an advantage over GUIFetch in the strict strategy.
By the relaxed strategy, our W-AE has comparable performance as its performance in the scaling-10\% and removal-20\% experiments (see Figure~\ref{fig:autoresults}).
In the moderate strategy, our W-AE still achieves precision@1=0.5 and MRR=0.62.
By the strict strategy, our W-AE remains a small advantage over GUIFetch.
Since this study involves many variations, it is hard to reach an agreement for five participants.
The GUIFetch baseline in reality no longer has the perfect performance as it does in the component-removal experiments.
Its performance is also much worse than that of some scaling experiments where the GUIFetch performs well.

\vspace{2mm}
\noindent\fbox{\begin{minipage}{13.5cm} \emph{Our CNN-based method can robustly retrieve relevant UI designs for a set of diverse, unseen real-application query UIs. In contrast, individual component-matching based heuristics find much fewer relevant UI designs for these real-application query UIs.} \end{minipage}}\\

\subsection{RQ4 Results: Usefulness Evaluation}

Finally, to answer RQ4 and evaluate the usefulness of our UI design search engine in real-world UI design tasks, where the users design and draw the UI wireframes on the fly during which they use our UI design search engine to search relevant UI designs.

\begin{figure*}
	\centering
	\includegraphics[width=0.4\textwidth]{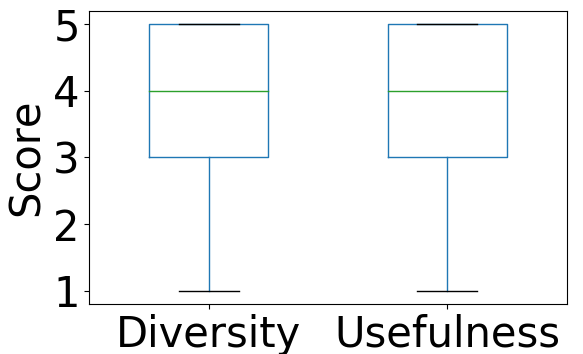}\
	%\vspace{-3mm}
	\caption{Boxplot for diversity and usefulness ratings by participants}
	%\vspace{-5mm}
	\label{fig:Quantitative results of usefulness evaluation}
\end{figure*}

\begin{figure*}
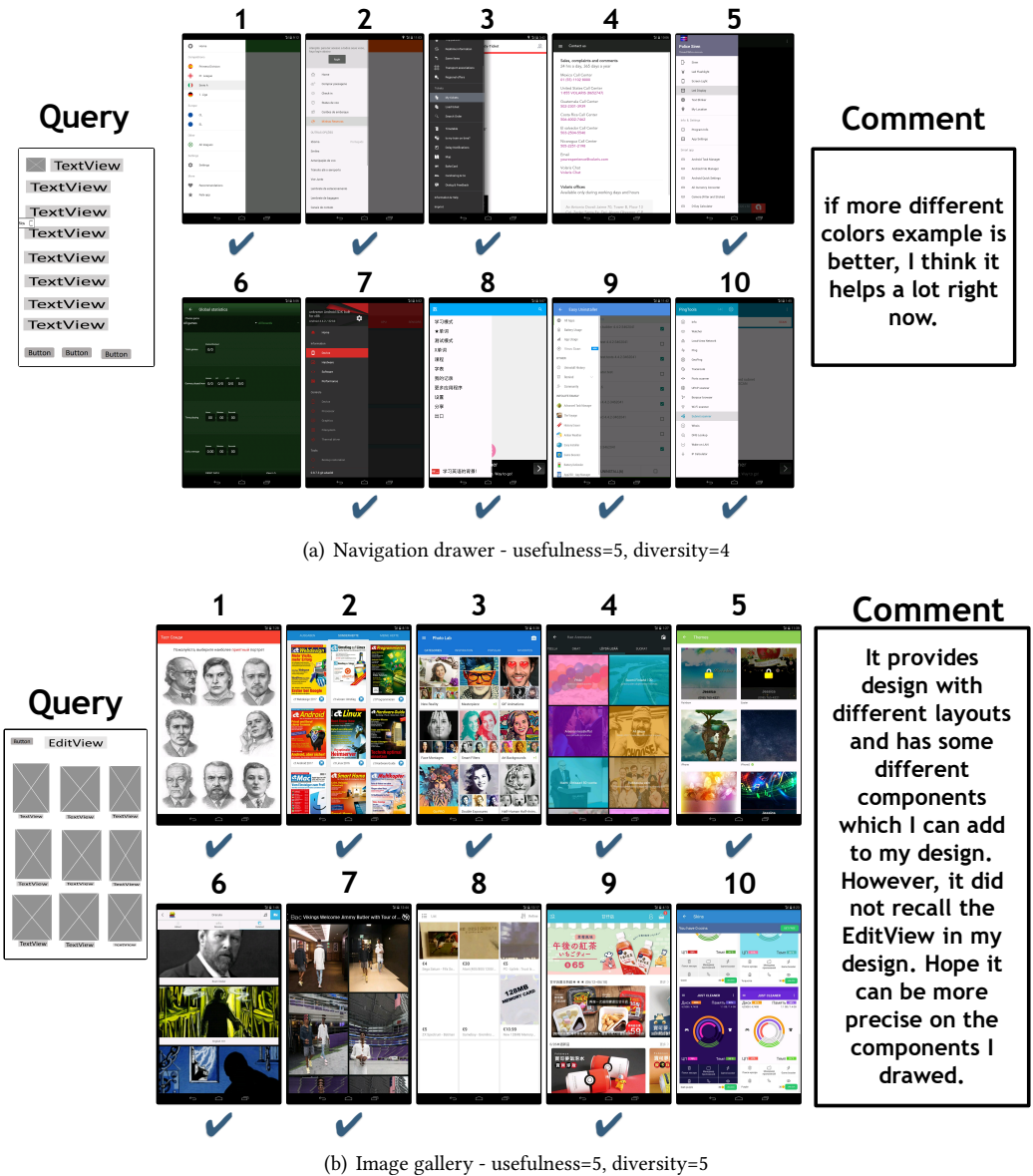

	\centering
	\subfigure[Navigation drawer - usefulness=5, diversity=4]{%
		\includegraphics[width=0.98\textwidth]{figures/usefulness/userstudy2/UserStudy2_navigation_drawer2.pdf}%
		\label{fig:Navigation drawer}%
	}
	\hfill
	\subfigure[Image gallery - usefulness=5, diversity=5]{%
		\includegraphics[width=1.0\textwidth]{figures/usefulness/userstudy2/UserStudy2_img_gallery2.pdf}%
		\label{fig:Image gallery}%
	}
	\caption{Examples of the search results in our user study (check marks indicate that users consider a design relevant)}
	\label{fig:results of userstudy2}
\end{figure*}

\subsubsection{Results}
%Participants
The 18 participants perform in total 168 times of search in the five UI design tasks.
Among the 90 participant-task sessions, 59 has one search, 12 has two searches, and 20 has three or more searches.
The times of search is reasonable considering the short experiment time for each task, as well as the time for drawing UI wireframes.
According to the participants' relevance judgment of search results, our W-AE achieves precision@1=0.44, precision@5=0.40, precision@10=0.38, and MRR=0.59.
These performance metrics fall in between those for the strict and relaxed strategies in our generation study, which demonstrates the practicality our search engine in support of real-world UI design tasks.

Figure~\ref{fig:Quantitative results of usefulness evaluation} shows the boxplot of diversity and usefulness ratings of the search results by the 18 participants.
For both these two aspects, the results from our model earn the scores of a median of 4 and the majority of them have a score falling in the range of 3 to 5, which indicates that our UI design search engine is satisfying.
Besides,among all 90 searches they rate, the participants rate the search results' diversity at 4 or 5 for 57 (63.3\%) searches, and rate the search results' usefulness at 4 or 5 for 51 (56.7\%) searches.
The motivating scenario illustrated in Section~\ref{sec:motivation} is actually derived from the design work by one participant in our user study.
We can observe the diversity and usefulness of the search results for inspiring that participant's design of sign-up UI.
Figure~\ref{fig:results of userstudy2} shows two more examples of the search results for the design of navigation drawer and image gallery respectively.
For the two user-drawn UI wireframes, our tool returns many relevant UI designs as annotated by the users (highlighted in blue check).
Furthermore, the users give 4 or 5 ratings for the diversity and usefulness of the search results, and provide some positive feebacks and useful suggestions on the search results.
Even for the irrelevant UI design (e.g., the 4th UI for the query navigation drawer wireframe), our model's recommendation stills makes some sense as that UI is visually similar to the query wireframe.
For the image gallery search results, in addition to the top-3 UI designs that have almost the same UI layout as the query UI wireframe, the other returned UI designs demonstrates diverse UI layouts for designing image gallery.

Our search engine does not produce satisfactory search results for 17 searches according to the participants' 1 or 2 diversity and usefulness ratings.
By interviewing the participants, we identify two main reasons for unsatisfactory search results.
First, our model tends to return the UI designs that are overall similar to the query UIs.
Although this improves the diversity of the search results which is beneficial for gaining design inspirations, it cannot guarantee the presence of some particular UI components or a particular component layout in the returned UI designs that the users want (see the feedback on the search results for image gallery in Figure~\ref{fig:results of userstudy2}).
To solve this problem, we may consider more advanced model such as variational autoencoder~\cite{doersch2016tutorial} which can force a greater loss when some user-desired components or component layouts in the query UI do not appear in the search results.

Second, some participants complain that our model are sometimes strict to the location of the components in a UI.
For example, when the user draws the switch buttons in the middle region of a preference setting UI, our tool does not return relevant UI designs.
But when he moves the switch button to the right side of the UI, our tool can return many relevant preference setting UIs.
This example actually shows that our model learns very well the characteristics of preference setting UIs in which switch buttons usually appear on the right side of the UI.
Although this modeling capability is desirable to filter out irrelevant UIs, it may make the search of relevant UIs too strict to a particular component layout.
To relax the search results, we may use structure similarity of images~\cite{wang2004image} or attribute graph~\cite{prabhu2015attribute} which support more abstract encoding of the component layout in UIs, and thus more flexible UI design search.

\section{Threats to Validity}
\label{sec:threatsToValidity}

We discuss two types of threats of validity in our work, namely, internal validity and external validity.

\subsection{Internal Validity}
Internal validity refers to the threat that may impact the results to causality~\cite{wohlin2012experimentation}.
First, our automatic evaluation allows us to conduct large-scale experiments to understand our approach's strengths and weaknesses, but it considers only component-scaling and component-removal variations separately.
Real-world UI design variations would be much more complex.
However, in order to dive into the influence that each treatment brings, we need to control the variable and it is also not feasible to try every combination of these two treatments.
To alleviate this influence, we further conduct generalization experiments and a user study to evaluate our tool by human participants.
The performance of our approach in these studies aligns well with that of our automatic evaluation, which gives us confidence in the practicality of our approach for real-world UI design search.

Second, to confirm the generalization of our model, we recruited five students with over two years of experience in Android development to manually examine the results from our model and baselines.
However, the notion of the concept of relevance may vary among them and thus impact the results.
Some of them may put more emphasis on the semantic meaning of the UI, while some may consider a UI comprised of similar components as relevant.
They both make sense since designers may directly reuse the design from the same scenario, but also get inspirations from UIs with similar layout and similar constructions.
To keep the evaluation consistent among different participants, we gave them a tutorial to learn the general meaning of these concepts, and a 15-minute warm-up time to get familiar with the tool and the experimental process.
Besides, the Cohen's kappa values and the Fleiss's kappa value indicate a moderate agreement between these participants.
It is reasonable since the variations we stated above.
We involved five students to try and avoid potential bias as best we can and analyse the results in terms of three strategies, namely strict, moderate and relaxed strategies, as states in Section~\ref{sec:rq3_metrics}.
We assert that by involving five participants and analysing results in terms of these three strategies, this threat to validity is reasonably mitigated.
Albeit participants' variance, the overall results still show that our approach outperforms other baselines by aggregating their feedback.

\subsection{External Validity}

External validity refers to the threat that may limit the ability to generalize~\cite{wohlin2012experimentation}.
First, our data collection tool could collect the majority UI elements from application, but could not capture the detailed HTML elements in the WebView component and some elements in UIs which require some specific engines, such unity3d game engines.
Therefore, such limitation may make us lose the UI designs from web components and game UIs.
However, the GUI design in these UI which contain HTML elements should be like those UIs which use the native elements.
The different implementation is merely an alternative to construct the user interface, while the underlying design principles should be the same.
We collect a database of 54,987 UI screenshots from 25 categories of 7,746 top-downloaded Android applications and we believe such large-scale database could cover the majority of UI designs.
We let the extension of our tool to collecting UI elements in WebView components and in specific engine as the future work.
Second, our approach is general, but our current tool supports only Android UI design search.
To further validate the generalizability of our approach, not only should the current tool be further enriched with more UI designs from further Android applications, but it should also be extended to other applications (e.g., iOS, web application).
Currently, we have already tested our model/tool on 25 categories of Android apps which demonstrates the generalization of our tool to some extent.
As the composition of a user interface is similar in terms of these platforms, we believe that our approach can also be applied with some customization.
But this will need to be further explored in the future.

\section{Related Work}
\label{sec:relatedWork}

\textbf{UI design datasets.}
Many UI design kits~\cite{web:UIkit1, web:UIkit2, web:UIkit3, web:UIkit4} are publicly available on the Web.
Designers also share their UI designs on social media platforms such as Dribbble~\cite{web:dribbble}, UI Movement~\cite{web:UImovement}.
They are a great source for design inspirations, but they cannot expose developers to a large UI design space of real applications.
Furthermore, these platforms support only simple keyword-based search.
Alternatively, existing applications provide a large repository of UI designs.
To harness these UI designs, people resort to automatic GUI exploration methods to simulate the user interaction with GUI and collect UI screenshots of existing applications, which can support data-driven applications such as UI code generation~\cite{nguyen2015reverse, chen2018ui, moran2018machine}, GUI search~\cite{behrang2018guifetch,bernal2019guigle, ritchie2011d, chen2019gallery}, design mining~\cite{kumar2013webzeitgeist}, design linting~\cite{zhao2020seenomaly}, UI accessibility~\cite{chen2020unblind}, user interaction modeling~\cite{deka2017rico, chen2019storydroid} and privacy and security~\cite{chen2019gui}.
In the same vein, our work builds a large database of real-application UI designs by automatic GUI exploration.
Different from existing work, we further wirify UI screenshots to support wireframe-based UI design search.

\textbf{UI design search.}
Our method allows users to search UI designs by UI wireframe images.
Some techniques~\cite{bernal2019guigle, ritchie2011d, kumar2013webzeitgeist} also support UI search by images, but they use low-level image features such as color histogram together with other UI information (if available) such as component type,  text displayed.
Other techniques~\cite{reiss2018seeking, behrang2018guifetch, zheng2019faceoff} support GUI search by UI sketches.
But they essentially convert both query UI and UIs in the database into a tree of GUI components and then find similar GUIs by computing the optimal matching of component trees.
Different from these works, our approach truly models UIs as images and uses deep learning features to encode the visual semantics of UIs.
The most related work is Rico~\cite{deka2017rico}, which envisions the possibility of deep learning based UI search and demonstrates several examples based on simple fully-connected layers model with highly simplified data.
Compared with their work, we develop a sophisticated model suitable for the variety of real-life UI designs, implement a working prototype and conduct systematic empirical studies.

\textbf{UI implementation automation.}
Nguyen and Csallner~\cite{nguyen2015reverse} detect components in UI screenshots by rule-based image processing method and generate GUI code.
They support only a small set of most commonly used GUI components.
More powerful deep-learning based methods~\cite{chen2018ui, beltramelli2018pix2code, moran2018machine} have been recently proposed to leverage the big data of automatically collected UI screenshots and corresponding code.
Different from these UI code generation methods which require high-fidelity UI design image, our approach requires only UI wireframes which can be fast prototyped even for inexperienced developers.
Furthermore, our method returns a set of diverse UI designs for exploring the design space, rather than the code implementing a specific UI design.
Some recent works explore issues between UI designs and their implementations.
Moran et al.~\cite{moran2018automated} check if the implemented GUI violates the original UI design by comparing the images similarity with computer vision techniques.
A follow-up work by them~\cite{moran2018detecting} further detects and summarizes GUI changes in evolving mobile apps.
UI design search finds similar UI designs, and then these techniques may be applied to further detect the differences between similar UI designs which may help refine the search results.

\section{Conclusion}
\label{sec:conclusion}
This paper presents a novel deep learning based approach for UI design search.
At the core of our approach is a UI wireframe image autoencoder.
Adopting image autoencoder architecture removes the barrier, i.e., labeled relevant UI designs which is impossible to prepare at large scale, for training UI design encoder.
Trained using a large database of unlabeled UI wireframes automatically-collected from existing applications, our wireframe encoder learns to encode more abstract and richer visual semantics of the whole UI designs than keywords, low-level image features and component type/position/size matching heuristics, leading to superior performance than the search methods based on these types of primitive information.
Our approach demonstrates the promising usefulness in supporting developers to explore and learn about a large UI design space.
As the first technique of its kind, our empirical studies also reveal technical and user needs for developing more robust and more usable UI design search methods.

\begin{acks}
This research was partially supported by the Australian National Univeristy - Data 61 Collaborative Researh Project (CO19314), the Australian Research Council's Discovery Early Career Researcher Award (DECRA) funding scheme (DE200100021), ARC Discovery Project scheme (DP170101932) and Laureate Fellowship (FL190100035).
\end{acks}

\bibliographystyle{ACM-Reference-Format}
\bibliography{reference}

%\appendix

%%\section{Research Methods}


\end{document}